\documentclass[preprint,showpacs,12pt,floatfix,aps]{revtex4}
\usepackage{graphicx}
\usepackage{latexsym}
\usepackage{amsmath}
\input psfig.sty

\newcommand{\myfigure}[3]{
        \begin{figure}
        \centerline{
        \includegraphics{#1.eps}}
        \caption{#2}
        \label{#3}
        \end{figure}
}

\newcommand{\bq}{\begin{equation}}
\newcommand{\eq}{\end{equation}}
\newcommand{\bqn}{\begin{eqnarray}}
\newcommand{\eqn}{\end{eqnarray}}
\newcommand{\nb}{\nonumber}
\newcommand{\lb}{\label}
\newcommand{\rr}{\bf r}

\begin{document}
\title{How the Cosmological Constant Affects Gravastar Formation}
\author{R. Chan $^{1}$}
\email{chan@on.br}
\author{M.F.A. da Silva $^{2}$}
\email{mfasnic@gmail.com}
\author{P. Rocha $^{34}$}
\email{pedrosennarocha@gmail.com}
\affiliation{$^{1}$ Coordena\c{c}\~ao de Astronomia e Astrof\'{\i}sica, Observat\'orio
Nacional, Rua General Jos\'e Cristino, 77, S\~ao Crist\'ov\~ao, CEP 20921-400,
Rio
de Janeiro, RJ, Brazil\\
$^{2}$ Departamento de F\'{\i}sica Te\'orica,
Instituto de F\'{\i}sica, Universidade do Estado do Rio de Janeiro,
Rua S\~ao Francisco Xavier 524, Maracan\~a,
CEP 20550-900, Rio de Janeiro - RJ, Brasil\\
$^{3}$ Instituto de F\'{\i}sica, Universidade Federal Fluminense, 
Av. Litor\^anea s/n, Boa Viagem, CEP 24210-340, Niter\'oi, RJ, Brazil\\
$^4$ Ger\^encia de Tecnologia da Informa\c c\~ao, ACERP, TV Brasil, 
R\'adios Nacional e MEC,
Rua da Rela\c c\~ao 18,  Lapa, CEP 20231-110, Rio de Janeiro, RJ, Brazil}
 
\date{\today}

\begin{abstract}
Here we generalized a previous model of gravastar consisted of an internal 
de Sitter spacetime, a dynamical infinitely thin  shell with
an equation of state, but now we consider an external 
de Sitter-Schwarzschild spacetime.
We have shown explicitly that the final output can be a black
hole, a "bounded excursion" stable gravastar, a stable gravastar, or a de Sitter 
spacetime, depending on the total mass of the system, the cosmological 
constants, the equation of state of the thin shell and
the initial position of the dynamical shell.
We have found that the exterior cosmological constant imposes a limit
to the gravastar formation, i.e., the exterior cosmological constant must be 
smaller than the interior cosmological constant. 
Besides, we have also shown that, in the particular case
where the Schwarzschild mass vanishes, no stable gravastar can be formed, 
but we still have formation of black hole.
\end{abstract}

\pacs{98.80.-k,04.20.Cv,04.70.Dy}

\maketitle

\section{Introduction}

As alternatives to black holes,  gravastars have received some attention
recently \cite{grava1}\cite{grava2}, partially due to the tight connection between the 
cosmological constant and a currently accelerating universe \cite{DEs}, 
although very strict observational constraints on the existence of such 
stars may exist \cite{BN07}. 

The pioneer model of gravastar was proposed by Mazur and Mottola (MM) \cite{MM01}.  After this work,
Visser and Wiltshire (VW) \cite{VW04} pointed out that there are two different types of stable 
gravastars which are stable gravastars and "bounded excursion" gravastars.
The first one represents a stable structure already formed, while the second one
is a system with a shell which oscillates around a equilibrium position which can loose energy and
to stabilize at the end. 

Recently we have done an extensive study on the problem of the stability of gravastars.
The first model \cite{JCAP} consisted of an internal de Sitter spacetime, a dynamical infinitely thin  shell of
stiff fluid, and an external Schwarzschild spacetime, as proposed by VW \cite{VW04}.
We have shown explicitly that the final output can be a black
hole, a "bounded excursion" stable gravastar, a Minkowski, or a de Sitter spacetime,
depending on the total mass $m$ of the system,  the cosmological constant $\Lambda$, and
the initial position $R_{0}$ of the dynamical shell. 
Therefore, we have shown, for the first time in the literature, that although it does
exist a region of the space of the initial parameters where it is always formed
stable gravastars, it still exists a large region of this space where
we can find black hole formation.  Then, we conclude that gravastar is not
an alternative model to black hole as it was originally proposed by VW models \cite{VW04}.

In the second paper \cite{JCAP1}, we have generalized the previous work on the problem of
stable gravastars considering an equation of state $p = (1-\gamma)\sigma$ for
the shell, instead of only using a stiff fluid ($\gamma=0$).
We have found that stable gravastars can be formed even for $\gamma \ne 0$,
since $\gamma < 1$, generalizing the gravastar models proposed until now. 
We also have confirmed the previous results, i.e., that both gravastars and black
holes can be formed, depending on the initial parameters.  

In the third work \cite{JCAP2}, we have generalized the former one considering now
an interior constituted by an anisotropic dark energy fluid.
We have again confirmed the previous results, i.e., that both gravastars and black
holes can be formed, depending on the initial parameters.  
It is remarkable that for this case we have an interior fulfilled by a physical matter,
instead of a de Sitter vacuum.  Thus, it is similar to phantom energy star models.

Nowadays, several kinds of observational data indicate that
our universe is in accelerated expansion. 
In Einstein's general relativity, in order to have such an
acceleration, one needs to introduce a component to the matter
distribution of the universe with a large negative pressure. This
component is usually referred as  dark energy. Astronomical
observations indicate that our universe is flat and currently
consists of approximately $2/3$ dark energy and $1/3$ dark
matter. The nature of dark energy as well as dark matter is
unknown, and many radically different models have been proposed,
such as, a tiny positive cosmological constant.  Based on this fact,
we would like to ask how the picture of the evolution
of gravastar formation is influenced by an exterior spacetime with
a positive cosmological constant.

Recently, Carter \cite{Carter} studied spherically symmetric gravastar 
solutions which possess an
(anti) de Sitter interior and a (anti) de Sitter-Schwarzschild or
Reissner-Nordstrom exterior. He followed the same approach that Visser and
Wiltshire took in their work \cite{VW04} assuming a potential $V(a)$ and
then founding the equation of state of the shell. He found a wide range
of parameters which allows stable gravastar solutions, and presented the
different qualitative behaviors of the equation of state for these
parameters.

Differently from Carter's work \cite{Carter}, we consider here another approach.
We generalize our second work in gravastars \cite{JCAP1}, introducing an
external de Sitter-Schwarzschild spacetime, to study how the cosmological constant
affects the gravastar formation.
We first assumed an equation of state, $p=(1-\gamma)\rho$, and, using
Israel conditions, derived a potential depending on the parameters of the
interior, the shell and the exterior of the gravastar's prototype.
We, then, studied the types of compact objects that can be generated
according to this potential, to the parameters related to the
cosmological constants and to the masses of our model. We found that both
gravastars and black holes can be formed.

The paper is  organized as follows: In Sec. II we present the metrics of the 
interior and exterior spacetimes, with theirs extrinsic curvatures,
the equation of motion of the shell and the potential of the system.
In Sec. III we discuss the particular cases where the Schwarzschild mass is null, 
and another where we have the same cosmological constant in the interior
and the exterior of the thin shell, which is presented in section IV.
In Sec. V we investigate the formation of
gravastar from numerical analysis of the general potential. 
Finally, in Sec. VI we present our conclusions.

\section{Formation of Gravastars in a de Sitter-Schwarzschild spacetime}

The interior spacetime is described by the de Sitter metric given by 
\bq
ds^2_{i}=-f_1 dt^2 + f_2 dr^2 + r^2 d\Omega^2,
\lb{ds2-}
\eq
where $f_1=1- (r/L_i)^2$,
$f_2= \frac{1}{1 - (ri/L_i)^2}$, $L_i=\sqrt{3/\Lambda_i}$ and 
$d\Omega^2 = d\theta^2 + \sin^2(\theta)d\phi^2$.

The exterior spacetime is given by a de Sitter-Schwarzschild metric
\bq
ds^2_{e}= - f dv^2 + f^{-1} d{\rr}^2 + {\rr}^2 d\Omega^2,
\lb{ds2+}
\eq
where $f=1 - \frac{2m}{\rr}-({\rr}/L_e)^2$ and $L_e=\sqrt{3/\Lambda_e}$.

The metric of the hypersurface do the shell is given by
\bq
ds^2_{\Sigma}= -d\tau^2 + R^2(\tau) d\Omega^2.
\lb{ds2Sigma}
\eq

Since $ds^2_{i} = ds^2_{e} = ds^2_{\Sigma}$ then $r_{\Sigma}={\rr}_{\Sigma}=R$,
and besides
\bqn
\dot t^2&=&\left[ f_1 - f_2 \left( \frac{\dot R}{\dot t} \right)^2 \right]^{-1}= \nb \\
& & \left[ 1- \left(\frac{R}{L_i}\right)^2+ \dot R^2 \right] \left[1- \left(\frac{R}{L_i}\right)^2 \right]^{-2},
\lb{dott2}
\eqn
and
\bqn
\dot v^2&=&\left[ f - f^{-1} \left( \frac{\dot R}{\dot v} \right)^2 \right]^{-1}= \nb \\
& & \left[ 1- \frac{2m}{R}-\left(\frac{R}{L_e}\right)^2+ \dot R^2 \right] \left[1- \frac{2m}{R}-\left(\frac{R}{L_e}\right)^2 \right]^{-2},
\lb{dotv2}
\eqn
where the dot represents the differentiation with respect to $\tau$.

Thus, the interior and exterior normal vector are given by
\bq
n^{i}_{\alpha} = (-\dot R, \dot t, 0 , 0 ),
\lb{nalpha-}
\eq
and
\bq
n^{e}_{\alpha} = (-\dot R, \dot v, 0 , 0 ).
\lb{nalpha+}
\eq

The interior and exterior extrinsic curvature are given by
\bqn
K^{i}_{\tau\tau}&=&-[(3 L_i^4 \dot R^2-L_i^4 \dot t^2+2 L_i^2 R^2 \dot t^2-
R^4 \dot t^2) R \dot t-(L_i+R) (L_i-R) (\dot R \ddot t-\ddot R \dot t) L_i^4] \times \nb \\
& &(L_i+R)^{-1} (L_i-R)^{-1} L_i^{-4}
\lb{Ktautau-}
\eqn
\bq
K^{i}_{\theta\theta}=\dot t (L_i+R) (L_i-R) L_i^{-2} R
\lb{Kthetatheta-}
\eq
\bq
K^{i}_{\phi\phi}=K^{i}_{\theta\theta}\sin^2(\theta),
\lb{Kphiphi-}
\eq
\bqn
K^{e}_{\tau\tau}&=&\dot v [(2 L_e^2 m \dot v+L_e^2 R \dot R-L_e^2 R \dot v+R^ 
3 \dot v) (2 L_e^2 m \dot v-L_e^2 R \dot R-L_e^2 R \dot v+R^3 \dot v)- \nb \\
& &2 L_e^4 R^2 \dot R^2] ((2 m-R) L_e^2+R^ 3)^{-1} 
(L_e^2 m-R^3) L_e^{-4} R^{-3}+\dot R \ddot v- \ddot R \dot v
\lb{Ktautau+}
\eqn
\bq
K^{e}_{\theta\theta}= -\dot v((2 m-R) L_e^2+R^3) L_e^{-2}
\lb{Kthetatheta+}
\eq
\bq
K^{e}_{\phi\phi}=K^{e}_{\theta\theta}\sin^2(\theta).
\lb{Kphiphi+}
\eq

Since we have \cite{Lake}
\bq
[K_{\theta\theta}]= K^{e}_{\theta\theta}-K^{i}_{\theta\theta} = - M,
\lb{M}
\eq
where $M$ is the mass of the shell, thus
\bq
M=\dot v R\left[ 1- \frac{2m}{R}-\left(\frac{R}{L_e}\right)^2+ \dot R^2 \right]+\dot t R \left[1- \left(\frac{R}{L_i}\right)^2 \right].
\lb{M1}
\eq

Then, substituting equations (\ref{dott2}) and (\ref{dotv2}) into (\ref{M1}) 
we get
\bq
M+R\left[1-\frac{2m}{R} -\left(\frac{R}{L_e}\right)^2 + \dot R^2 \right]^{1/2} - 
R \left[ 1 -\left(\frac{R}{L_i}\right)^2  + \dot R^2 \right]^{1/2}=0.
\lb{M2}
\eq

Solving the equation (\ref{M2}) for $\dot R^2/2$ we obtain the potential $V(R,m,L_i,L_e)$.
In order to keep the ideas of our work \cite{JCAP1} as much as possible, we consider the thin 
shell as consisting
of a fluid with a equation of state, $\sigma = (1-\gamma)\vartheta$, where $\sigma$ and $\vartheta$ denote, 
respectively, the surface energy density and pressure of the shell and $\gamma$ is a constant. 
The equation of motion of the shell is given by \cite{Lake}
\bq
\dot M + 8\pi R \dot R \vartheta = 4 \pi R^2 [T_{\alpha\beta}u^{\alpha}n^{\beta}]=
\pi R^2 \left(T^+_{\alpha\beta}u_+^{\alpha}n_+^{\beta}-T^-_{\alpha\beta}u_-^{\alpha}n_-^{\beta} \right),
\lb{dotM}
\eq
where $u^{\alpha}$ is the four-velocity.  Since the interior 
and the exterior spacetimes correspond to vacuum solutions, we get
\bq
\dot M + 8\pi R \dot R (1-\gamma)\sigma = 0,
\lb{dotM1}
\eq
and since $\sigma = M/(4\pi R^2)$ we can solve equation (\ref{dotM1}) giving
\bq
M=k R^{2(\gamma-1)},
\lb{Mk}
\eq
where $k$ is an integration constant.

Substituting equation (\ref{Mk}) into $V(R,m,L_i,L_e)$ we obtain
\bqn
& &V(R,m,L_i,L_e,k,\gamma)= \nb \\
& &-\frac{1}{8R^2 L_e^4 L_i^4 k^2}
\left[-4 R^2 L_e^4 L_i^4 k^2+4 R m L_e^4 L_i^4 k^2+2 R^4 L_i^4 k^2 L_e^2 +R^{(-4 \gamma+12)} L_e^4 \right. \nb \\
& &\left. -4 R^{(-4 \gamma+9)} L_e^4 L_i^2 m-2 R^{(-4 \gamma+12)} L_e^2 L_i^2+2 R^4 L_e^4 L_i^2 k^2+4 R^{(-4 \gamma+6)} L_i^4 m^2 L_e^4 \right. \nb \\
& &\left. +4 R^{(-4 \gamma+9)} L_i^4 m L_e^2+R^{(-4 \gamma+12)} L_i^4+R^{(4 \gamma-4)} L_i^4 k^4 L_e^4 \right].
\eqn

Redefining the Schwarzschild mass $m$, the cosmological constants $L_i$ and $L_e$ and the radius $R$ as
\bq
m \equiv mk^{-\frac{1}{2\gamma-3}},
\eq
\bq
L_i \equiv L_i k^{\frac{2}{2\gamma-3}},
\eq
\bq
L_e \equiv L_e k^{\frac{2}{2\gamma-3}},
\eq
\bq
R \equiv Rk^{-\frac{1}{2\gamma-3}},
\eq
we get the potential
\bqn
& &V(R,m,L_i,L_e,\gamma)= \nb \\
& &-\frac{1}{2}\left[-1 + \frac{m}{R} +\frac{R^{(4\gamma-6)}}{4}+m^2R^{(-4\gamma+4)}+\frac{R^2}{2L_i^2}-\frac{mR^{(-4\gamma+7)}}{L_i^2}+
\frac{R^{(-4\gamma+10)}}{4L_i^4}- \right. \nb \\
& & \left.  -\frac{R^{(-4\gamma+10)}}{2L_i^2L_e^2}+ 
\frac{R^2}{2L_e^2}+\frac{mR^{(-4\gamma+7)}}{L_e^2}+
\frac{R^{(-4\gamma+10)}}{4L_e^4} \right].
\lb{VR}
\eqn

Redefining $L_e=\alpha L_i$ we finally get
\bqn
& &V(R,m,\alpha,L_i,\gamma)= \nb \\
& &-\frac{1}{2}\left[-1 + \frac{m}{R} +\frac{R^{(4\gamma-6)}}{4}+m^2R^{(-4\gamma+4)}+\frac{R^2}{2L_i^2}-\frac{mR^{(-4\gamma+7)}}{L_i^2}+
\frac{R^{(-4\gamma+10)}}{4L_i^4}- \right. \nb \\
& & \left.  -\frac{R^{(-4\gamma+10)}}{2\alpha^2L_i^4}+
\frac{R^2}{2\alpha^2 L_i^2}+\frac{mR^{(-4\gamma+7)}}{\alpha^2 L_i^2}+
\frac{R^{(-4\gamma+10)}}{4\alpha^4 L_i^4} \right].
\lb{VRa}
\eqn
It is curious to note that this potential is independent of the
sign of the parameter $\alpha$.

Therefore, for any given constants $m$, $\alpha$, $L_i$ and $\gamma$, equations (\ref{VR}) or (\ref{VRa}) uniquely determines the collapse
of the prototype  gravastar. Depending on the initial value $R_{0}$,  the collapse can
form either a black hole, or gravastar,  or a de Sitter spacetime.
In the last case, the thin shell
first collapses to a finite non-zero minimal radius and then expands to infinity.  To  guarantee
that initially the spacetime does not have any kind of horizons,  cosmological or event,
we must restrict $R_{0}$ to the ranges simultaneously,
\bq
\lb{2.2b}
2m<R_{0} < L_i,
\eq
\bq
2m<R_{0} < L_e,
\eq
where $R_0$ is the initial collapse radius.

In order to fulfill the energy condition $\sigma+2p\ge0$ of the shell
and assuming that 
$p=(1-\gamma)\sigma$ we must have $\gamma \le 1.5$. On the other hand, in order
to satisfy the condition $\sigma+p\ge 0$, we get that $\gamma \le 2$.
The dominant energy condition is only satisfied for $0 \le \gamma \le 2$.
Although the phantom energy is usually considered as a kind of dark energy,
in this paper we will use the expression dark energy for the case where the
condition $\sigma + p \ge 0$ is satisfied and phantom energy otherwise.
Hereinafter, we will use only some particular values of the parameter 
$\gamma$ which are analyzed in this work. See Table I.

Since the potential, equations (\ref{VR}) or (\ref{VRa}), is very complex to manipulate analytically,
we have analyzed several special cases.

\begin{table}
\caption{\label{tab:table1}This table summarizes the matter classification
based on the energy conditions of the shell, in terms of the parameter $\gamma$.}
\begin{ruledtabular}
\begin{tabular}{cccc}
Matter & Condition 1 & Condition 2  & $\gamma$ \\
\hline
Standard Energy           & $\sigma+2p\ge 0$ & $\sigma+p\ge 0$ & $\gamma \le 1.5$  \\
Dark Energy               & $\sigma+2p\le 0$ & $\sigma+p\ge 0$ & $1.5 \le \gamma \le 2$ \\
Phantom Energy  & $\sigma+2p\le 0$ & $\sigma+p\le 0$ & $\gamma \ge 2$ \\
\end{tabular}
\end{ruledtabular}
\end{table}

\section{Case $m=0$}

This case represents a system where the Schwarzschild mass vanishes and the
combination of both cosmological constant (interior and exterior) imposes
a very special junction thin shell.  Note that from equation (\ref{M2}),
this configuration is possible only if $\alpha \ne 1$, otherwise if
$\alpha=1$ then we have $M=0$, i.e., the thin shell vanishes.

From the equation (\ref{VRa}) we get
\bq
\frac{1}{2} - \frac{R^{(4\gamma-6)}}{8}-\frac{R^2}{4L_i^2}- 
\frac{R^{(-4\gamma+10)}}{8L_i^4}+ 
\frac{R^{(-4\gamma+10)}}{4\alpha^2L_i^4}- 
\frac{R^2}{4\alpha^2 L_i^2}- \frac{R^{(-4\gamma+10)}}{8 \alpha^4 L_i^4}=0 
\lb{V1}
\eq
and differentiating the potential $dV(R)/dR$, we get
\bqn
& &-\frac{R^{(4 \gamma-6)} (4 \gamma-6)}{8 R}-\frac{R}{2 L_i^2}-
\frac{R^{(-4 \gamma+10)} (-4 \gamma+10)}{8 R L_i^4}+
\frac{R^{(-4 \gamma+10)} (-4 \gamma+10)}{4 R \alpha^2 L_i^4}- \nb \\
& &\frac{R}{2 \alpha^2 L_i^2}-\frac{R^{(-4 \gamma+10)} (-4 \gamma+10)}{8 R\alpha^4 L_i^4}=0.
\lb{V2}
\eqn
From these two equations we can obtain the point where the potential
has a minimum and equal to zero.  Solving simultaneously the equations
(\ref{V1}) and (\ref{V2}) we get
\bqn
& &R_c=2^{-\frac{1}{2 \gamma-3}}\left\{-\frac{2}{\alpha^2 (-2+\gamma)} \left\{-2\alpha^4+\alpha^4 \gamma-2 \alpha^2 \gamma+6 \alpha^2+\gamma-2+ \right. \right. \nb \\
& &\left. \left. \left[(\alpha^2+1)^2 (4 \alpha^4+\alpha^4 \gamma^2-4 \alpha^4 \gamma-
7 \alpha^2-2 \alpha^2 \gamma^2+8 \alpha^2 \gamma+4+\gamma^2-4 \gamma)\right]^{\frac{1}{2}}\right\}
\right\}^\frac{1}{2 (2 \gamma-3)}, \nb \\
& &
\eqn
\bqn
& &L_c^i=-\frac{1}{(-4+R^{4 \gamma-6})\alpha} \left\{ -(-4+R^{4 \gamma-6}) \left[ R^2 \alpha^2+R^2+ \left( R^4 \alpha^4+2 R^4 \alpha^2+R^4+4 R^{(-4 \gamma+10)}- \nb \right. \right. \right.\\
& &\left. \left. \left. 8 R^{(-4 \gamma+10)} \alpha^2+4 R^{(-4 \gamma+10)} \alpha^4-R^{(4 \gamma-6)} R^{(-4 \gamma+10)}+2 R^{(4 \gamma-6)} R^{(4 \gamma+10)} \alpha^2-R^{(4 \gamma-6)} R^{(-4 \gamma+10)} \alpha^4 \right)^\frac{1}{2} \right] \right\}^{\frac{1}{2}}. \nb \\
& &
\eqn

For $\alpha=\infty$ we get the same results of previous work \cite{JCAP1}, given by
\bq
R_c  =  \left|\frac{2(\gamma - 2)}{2\gamma - 5}\right|^{\frac{1}{3 - 2\gamma}},
\eq
and
\bq
L_{c}^i = \left|\frac{2\gamma - 5}{2\gamma - 3}\right|^{1/2} R_{c}^{2(2-\gamma)}.
\eq

\myfigure{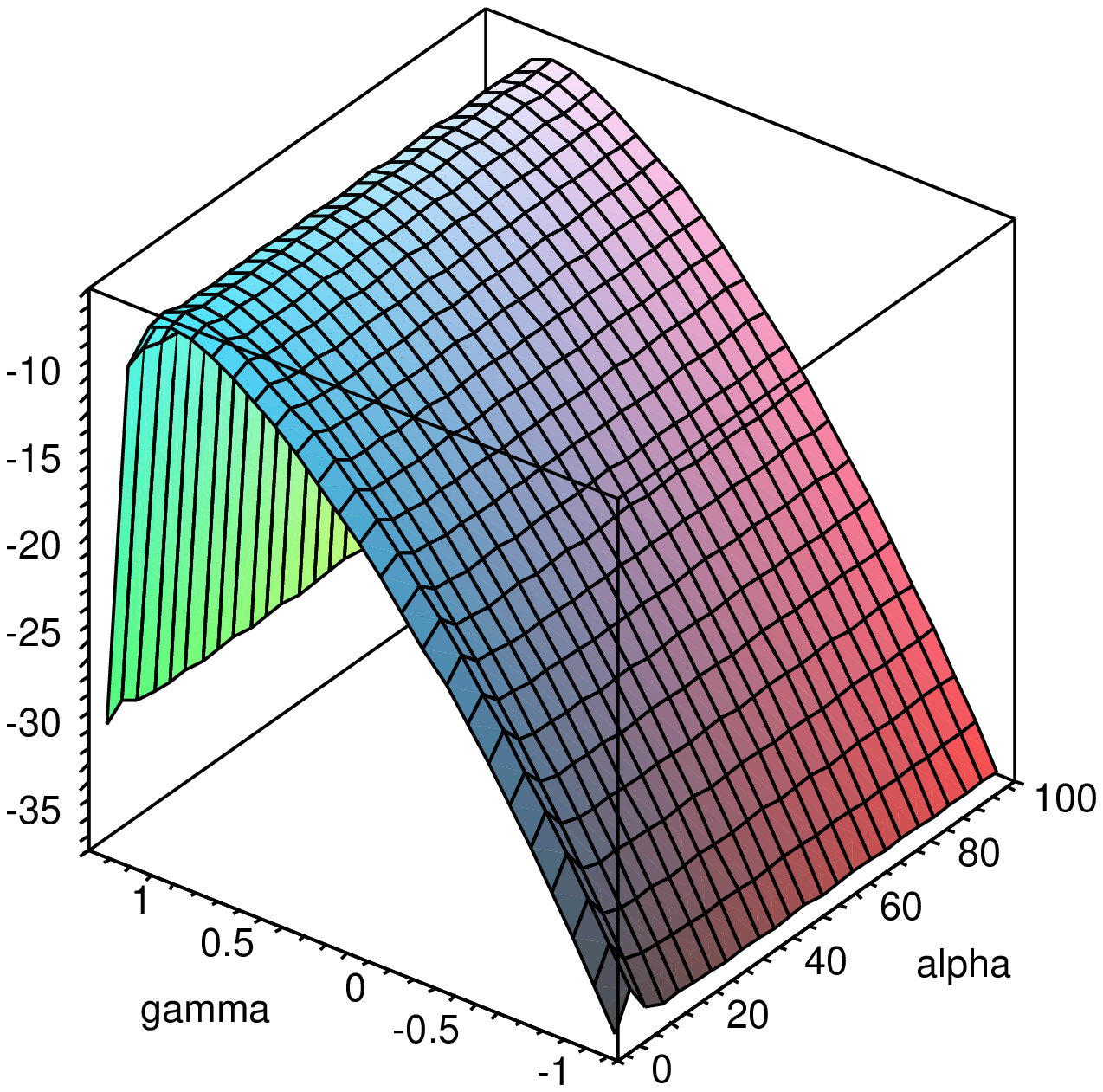}{This plot shows, in terms of $\alpha$ and $\gamma$,
 the second derivative of the potential
$V(R,m,\alpha,L_i,\gamma)$ with respect to $R$, calculated 
at the values $R=R_c$ and $L_i=L_c^i$, in the intervals $-1.5 <\gamma <1.5$ and
$0<\alpha<100$, for $m=0$.
}{test}

We can see from figure \ref{test} that the quantity $V"(R)$, calculated
at the values $R=R_c$ and $L_i=L_c^i$, is always negative,
for a large range of values for $\alpha$ and $\gamma$ ($-1<\gamma<1.5$ and
$0<\alpha<100$).  This means that, if we impose $V(R)=V'(R)=0$, we have always formation
of black holes, instead of formation of stable gravastars.

In the next sections, we will analyze another interesting particular case,
where $m \ne 0$ and $\alpha=1$.

\section{Case $m \ne 0$ and $\alpha=1$}

In this case we consider $L_e=L_i=L$ since $\alpha=1$.

From the equation (\ref{VRa}) we get
\bq
-1+{\frac {m}{R}}+\frac{1}{4}\,{R}^{4\,y-6}+{m}^{2}{R}^{-4\,y+4}+{\frac {{R}^{2}}{{L}^{2}}}=0
\lb{V1a}
\eq
and differentiating the potential $dV(R)/dR$, we get
\bq
-{\frac {m}{{R}^{2}}}+\frac{1}{4}\,{\frac {{R}^{4\,\gamma-6} \left( 4\,\gamma-6 \right) 
}{R}}+{\frac {{m}^{2}{R}^{-4\,\gamma+4} \left( -4\,\gamma+4 \right) }{R}}+2\,{
\frac {R}{{L}^{2}}}=0
\lb{V2a}
\eq
From these two equations we can obtain the point where the potential
has a minimum and equal to zero.  Solving simultaneously the equations
(\ref{V1a}) and (\ref{V2a}) we get
\bq
m_c= {\frac { \left( -3+\sqrt { \left( 25\,{R}^{4\,\gamma}+16\,{\gamma}^{2}{R}^{
4\,\gamma}-40\,\gamma{R}^{4\,\gamma}+32\,{R}^{6}\gamma-16\,{R}^{6} \right) {R}^{-4\,\gamma}}
 \right) {4R}^{4\,\gamma}}{{R}^{5} \left( 2\,\gamma-1 \right) }},
\eq
\bqn
& &L_c^2= \nb \\
& &{\frac {-8{R}^{13-8\,\gamma} \left( 2\,\gamma-1 \right) ^{2}}{16\,{R}^{11-8\,
\gamma}(3\gamma-2\gamma^2-1)+(4\gamma-5){R}^{-4\,\gamma+5}[4\gamma-5
+\sqrt {(4\gamma-5)^{2}+16\,{R}^{-4 \,\gamma+6}(2\gamma-1)}]}}. \nb \\
& &
\eqn

\myfigure{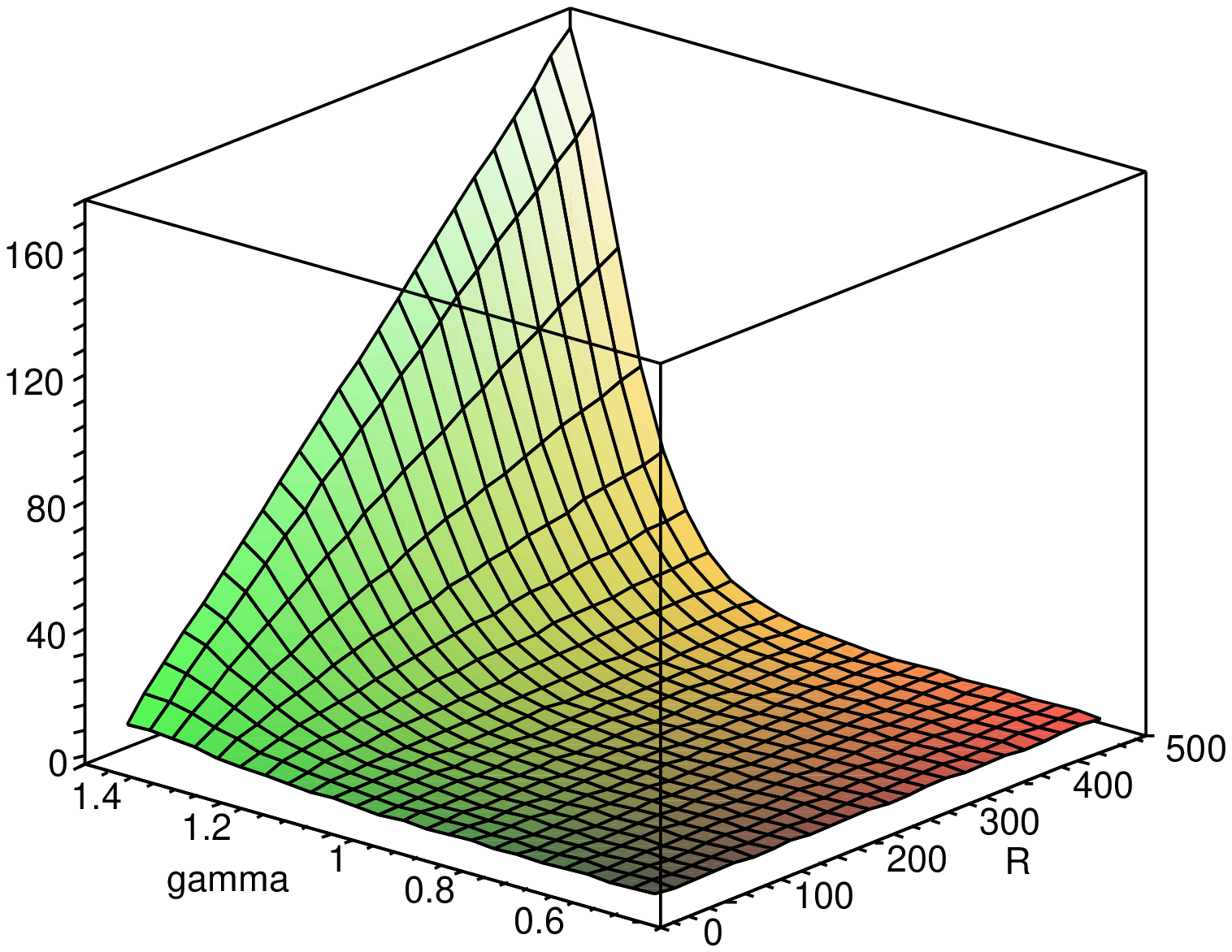}{This plot shows that, in terms of $R$ and $\gamma$ and for $L_e=L_i$,
the critical mass $m_c$ is always negative, when $V(R)=0$ and $V'(R)=0$, for the interval $0.5<\gamma<1.5$, 
since the critical mass is not defined for $\gamma<0.5$.}{mc}

\myfigure{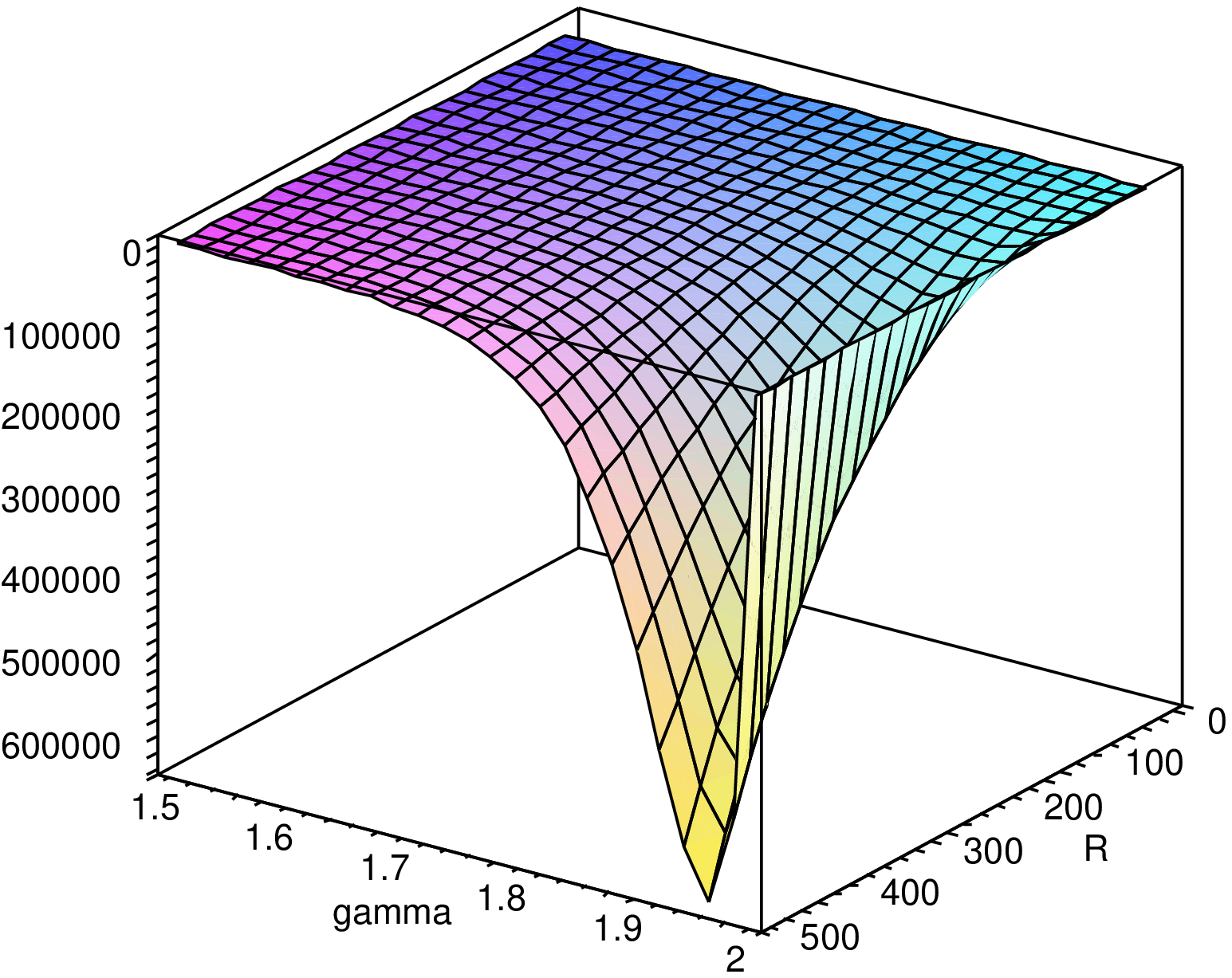}{This plot shows that, in terms of $R$ and $\gamma$ and for $L_e=L_i$,
the critical mass $m_c$ is always negative, when $V(R)=0$ and $V'(R)=0$, for the interval $1.5<\gamma<2$.
}{mc1}

\myfigure{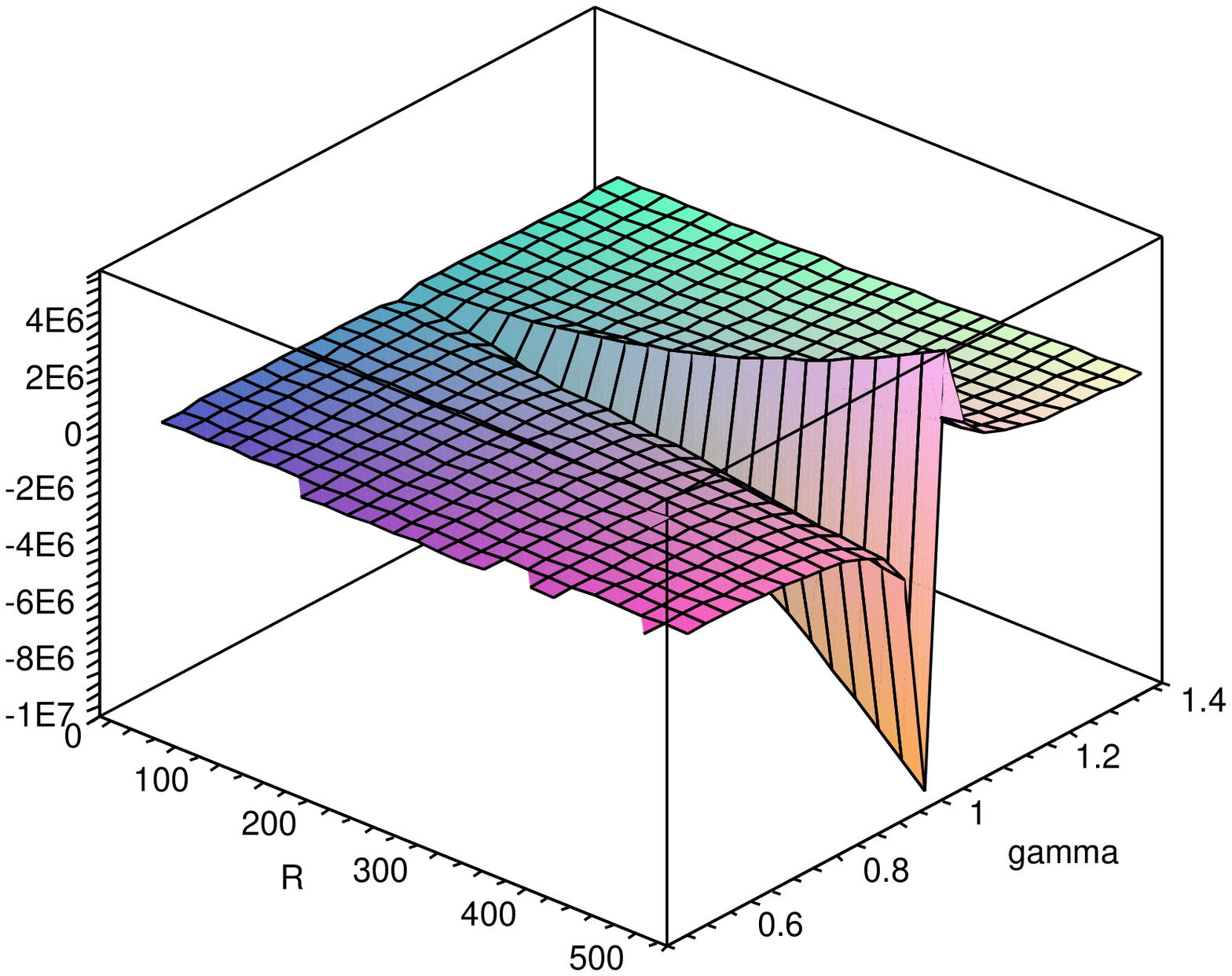}{This plot shows, in terms of $R$ and $\gamma$,
the cosmological constants $L_e=L_i$ when $V(R)=0$ and $V'(R)=0$, 
for the interval $0.5<\gamma<1.5$, since the cosmological constant is not
defined for $\gamma<0.5$.  The critical mass $m_c$ is always negative for 
$1.5 < \gamma < 2.0$.}{lc}

\myfigure{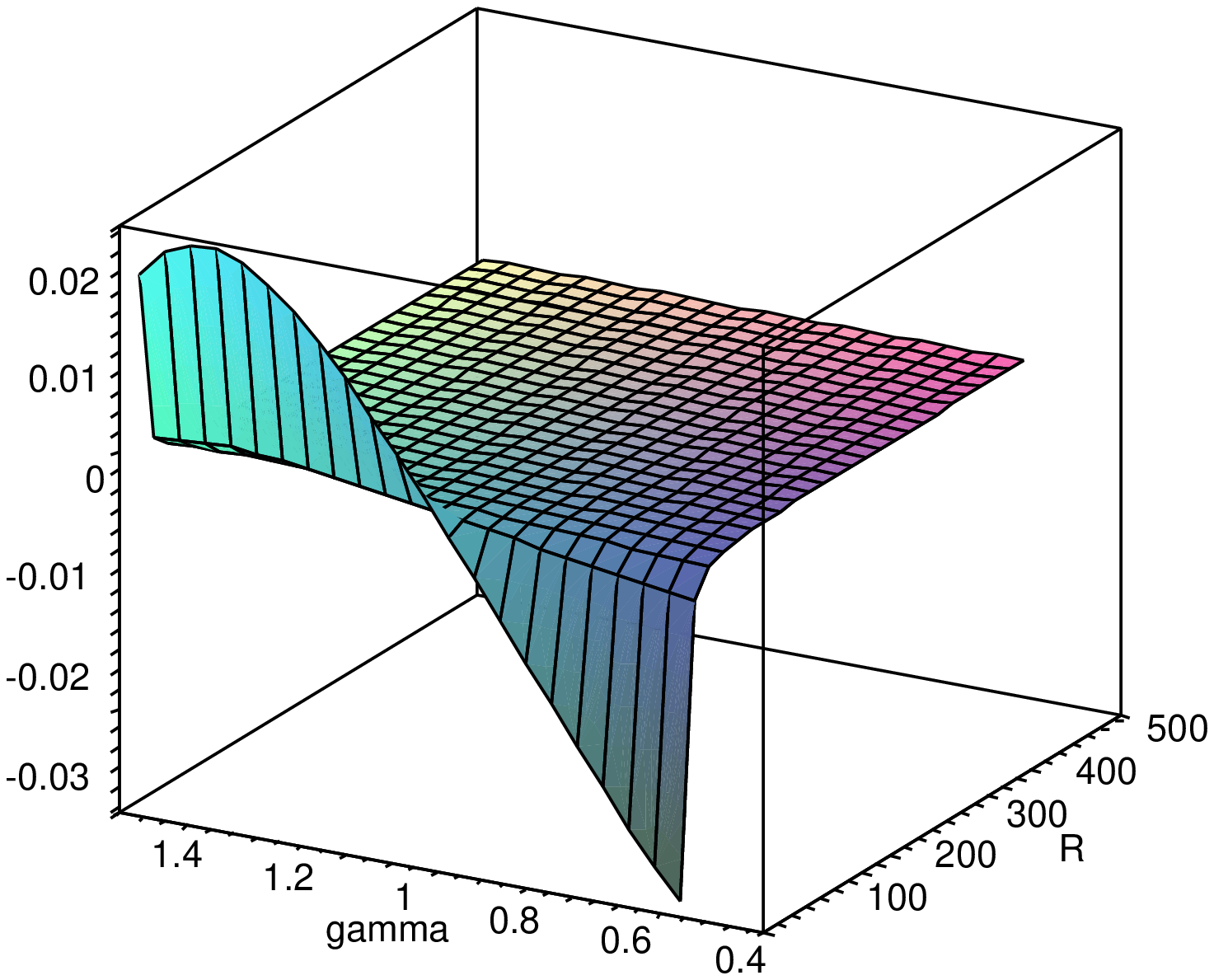}{This plot shows, in terms of $\alpha$ and $\gamma$,
the second derivative of the potential
$V(R,m,\alpha=1,L_i,\gamma)$ with respect to $R$, calculated 
at the values $R=R_c$ and $L_i=L_c^i$, for the interval $0.4<\gamma<1.5$,
since the derivative is not defined for $\gamma<0.5$.
}{diff2V}

From the figures \ref{mc} and \ref{mc1} we can see that the critical mass
$m_c$ is positive only in the range $0.5 \le \gamma \le 1.5$.  Besides,
from the figure \ref{lc} we can note that there is not any real value for
the critical cosmological constant $L_c$ in the interval $\gamma \le 1$.
As a consequence of these results, the second derivative of the potential
$V"(R)$, shown in the figure \ref{diff2V}, is negative for $\gamma < 1$ and
positive for $1 < \gamma \le 1.5$.
For $\gamma=1$ we have $V"(R)=0$ implying that we have an inflection point
in the potential.

Combining all these facts, we conclude that for $L_e=L_i$ we obtain the
following:

\begin{enumerate}

\item For $\gamma > 1.5$, which corresponds to a dark energy shell,
none structure is formed.

\item For $0.5 \le \gamma \le 1.5$, which corresponds to a standard fluid shell,
it can collapse to a black hole ($0.5 \le \gamma < 1$), or it does not
collapse, reaching an equilibrium stage, forming a stable gravastar 
($1 < \gamma \le 1.5$).

\item For $\gamma=1$ none gravastar is formed.

\end{enumerate}

Then, for $\alpha=1$, we have shown that no stable gravastar can exist, for
$\gamma \le 0.5$.  

\section{General Case}

The expressions for the potentials in the present case makes difficult a 
complete analytic analysis, so we shall study it numerically. Our main strategy 
is to start with the values of $m_{c}$ obtained for the case studied in our 
previous work \cite{JCAP1}, where $m \ne 0$ and $L_i \ne \infty$, and then 
gradually turn on $L_{e}$. The potential $V(R,m_{c},L_i,L_e,\gamma)$ 
is plotted as a function of $R$, by finely tuning $L_{e}$ until a stable gravastar 
or a "bounded excursion" gravastar is found. We also made another approach, solving 
the system of equations $V(R,m_{c},L_{i},L_{e},\gamma)= 0$ and 
$V'(R,m_{c},L_{i},L_{e},\gamma)=0$ for $R$ and $L_{e}$ and fixing the parameters 
$m_{c}$, $L_{i}$ and $\gamma$ in order to compare the results we obtained for $L_{e}$.  
It was seen that there is a range of $L_{e}$ in which "bounded excursion" stable 
gravastars are found, i.e.,  $L_{e}^{min} < L_e < L_{e}^{max}$. 
For $L_e \ge L_e^{max}$ we have found only stable gravastars.

We must call attention to the fact that, hereinafter, we will not consider the
physical situation where there is dispersion of the star.
If the initial radius of the collapse is greater enough, the star will first
contract to its minimal radius and then expand to infinity, whereby a de Sitter
spacetime is finally formed.

Figures \ref{figa}-\ref{figf} show the behavior of the potential as a function of 
$R$ for the case where $\gamma=-1$. This case was not studied on our previous 
work~\cite{JCAP1} where there was no cosmological constant external to the thin shell. 
So, we used the analytic expression (2.23) from our previous work~\cite{JCAP1} to 
calculate $m_{c}$. This situation is analogous to our present work if we use
the potential with $L_{i}=L_{e}=\infty$. The potential is shown in 
figure  \ref{figa} where $m_{c} \approx 0.5055981490$ and $R_{c} \approx 1.023836256$. 
For $m>m_{c}$ the potential $V(R)$ is strictly negative and the collapse always forms 
black holes. For $m=m_{c}$, there are two different possibilities, depending on the 
choice of the initial radius $R_{0}$. In particular, if the star begins to collapse 
with $R_{0}>R_{c}$, the collapse will asymptotically approach the minimal radius 
$R_{c}$. Once it collapses to this point, the shell will stop collapsing and remains 
there for ever. However, in this case this point is unstable and any small 
perturbations will lead the star either to expand for ever and leave behind a flat 
spacetime, or to collapse until $R=0$, whereby a Schwarzschild black hole is finally 
formed. On the other hand, if the star begins to collapse with $2m_{c}<R_{0}<R_{c}$, 
the star will collapse until a black hole is formed. For $m<m_{c}$, the potential 
$V(R)$ have a positive maximum, and the equation $V(R,m<m_{c})=0$ has two positive 
roots $R_{1,2}$ with $R_{2}>R_{1}>0$. There are two possibilities here, depending 
on the choice of the initial radius $R_{0}$. If $R_{0}>R_{2}$, the star will first 
collapse to its minimal radius $R=R_{2}$ and the expand to infinity, whereby a 
Minkowski spacetime is finally formed. If $2m_{c}<R_{0}<R_{1}$, the star will 
collapse continously until R=0, and a black hole will be finally formed. As we 
always have $V''<0$, it means that no stable stars exist in this case.

For the case of the figure \ref{figa}, i.e, $m=0.5055981490$, $L_i=L_e=\infty$
and $\gamma=-1$, we have analyzed the behavior of the potential for the
parameter $\gamma=0.7, 1.7, 3$ and we have found that we get only dispersion
of the shell.

The figures \ref{figb} and \ref{figc} show the case where $L_{e}=\infty$, but 
$L_{i} \ne \infty$. Variations of $m$ fixing the parameter $L_i$ and variations 
of $L_i$ fixing the parameter $m$ reveal that both stable gravastars and "bounded 
excursion" stable gravastars can be formed, but not excluding the existence 
of black holes.

For the general case, where both $L_{i}$ and $L_{e}$ are not infinity, it 
is shown the potential $V(R)$ as a function of $R$ for some specific values of 
$\gamma$, which are $\gamma=-1$, $\gamma=0$, $\gamma=0.4$ and $\gamma=0.7$ 
representing standard energy, $\gamma=1.7$ representing dark energy and
$\gamma=3$ for phantom energy. Note that $\gamma=-1$ and $\gamma=3$
violate the dominant energy condition.  Note also that in the Carter's work \cite{Carter}, the 
dominant energy condition is considered to restrict acceptable solutions.
In our case this corresponds to the cases  $\gamma=0$, 0.4, 0.7 and $\gamma=1.7$.
We found that the shell must have standard energy (figures \ref{figd}, \ref{fig1},
\ref{fig2} and \ref{fig3}) in order to have both stable gravastars or 
"bounded excursion" stable gravastars (the later existing whenever 
$L_{e}^{min}<L_e<L_{e}^{max}$ as explained in the text), but never excluding 
the existence of black holes or the formation of a de Sitter space depending 
on the choice of initial radius $R_{0}$ (It is important to verify 
the restriction on the values $R_{0}$ can assume, obeying both $2m<R_{0}<L_{i}$ 
and $2m<R_{0}<L_{e}$.). For dark energy shells and for phantom energy shells 
there are not formation of gravastars (figures \ref{fig4} and \ref{fig5}). 
Variations of $\gamma$ in the potentials studied also show that when the 
region of $\gamma$ represents dark or phantom energy, there are only possibilities 
of formation of black holes or de Sitter spacetime (figures \ref{figf}, \ref{fig7}, 
\ref{fig9} and \ref{fig12}).  When the shell is made of standard 
energy we can have gravastars or black holes 
(figures \ref{fige}, \ref{fig6}, \ref{fig8}, \ref{fig10} and \ref{fig11}).
Thus, we only find gravastars for standard energy shells, satisfying or not
the dominant energy conditions.

\begin{table}
\caption{This table summarizes all possible kind of energy
of the interior fluid and of the shell.  M, S, dS and dSS denote Minkowski,
Schwarzschild, de Sitter and de Sitter-Schwarzschild spacetimes, respectively.}
\begin{ruledtabular}
\begin{tabular}{ccccccc}
Case & Interior & Shell Energy & Exterior & Figures & Conditions & Structures \\
\hline
A &  M & Standard & S   & 6              & $m > m_{c}$                                           & Black Hole \\
B & dS & Standard & S   & 7              & $m=m_c$, $m=0.51$                                     & Gravastar \\
  &    &          &     & 7              & $m=0.53$                                              & Black Hole \\
  &    &          &     & 8              & $L_i=L_c$, $L_i=1.4$                                  & Gravastar \\
  &    &          &     & 8              & $L_i=1.0$                                             & Black Hole \\
  &    &	  & dSS & 9, 12, 13, 14  & ${L_e}^{mim}<L_e<{L_e}^{max}$                         & Gravastar \\
  &    &	  &     & 9, 12, 13, 14  & $L_e \le {L_e}^{min}$                                 & Black Hole \\
  &    &          &     & 10, 11         & $\gamma=-1$                                           & Gravastar \\
  &    &          &     & 17             & $\gamma=0.0$                                          & Gravastar \\
  &    &          &     & 19, 21         & $\gamma=0.4$                                          & Gravastar \\
  &    &          &     & 19, 21         & $\gamma=0.0$                                          & Black Hole \\
  &    &          &     & 22             & $\gamma=0.7$                                          & Gravastar \\
  &    &          &     & 22             & $\gamma=0.0, 0.4$                                     & Black Hole \\
C & dS & Dark     & dSS & 15             & $m>m_c$                                               & Black Hole \\
D & dS & Phantom  & dSS & 11, 18, 20, 23 & $\gamma=3$                                            & Black Hole \\
  &    &          &     & 16             & $m > m_c$                                             & Black Hole \\
\end{tabular}
\end{ruledtabular}
\end{table}

\section{Conclusions}

In this paper, we have generalized the problem of the stability of gravastars 
studied recently by us \cite{JCAP1}, introducing a positive cosmological
constant in the exterior spacetime. Thus, the model consists of a de Sitter
interior spacetime, a dynamical infinitely thin  shell of
fluid with an equation of state $p=(1-\gamma)\sigma$, and an external 
de Sitter-Schwarzschild spacetime.
We have shown explicitly that the final output can be a black
hole, a "bounded excursion" stable gravastar, a stable gravastar, or a de Sitter 
spacetime, depending on the total mass $m$ of the system, the parameter 
$\alpha$, 
the constant $L_i$, the parameter of the shell $\gamma$ and
the initial position $R_{0}$ of the dynamical shell. All these possibilities
have non-zero measurements in the parameter space of $m$, $L_i$, $\alpha$, $\gamma$ 
and $R_{0}$, for both gravastar and black hole. 

For $m=0$, the analysis of the potential has shown that, 
if we impose $V(R)=V'(R)=0$, we have always formation
of black holes, instead of formation of stable gravastars.
Comparing the results from \cite{JCAP1} ($L_e=\infty$) 
with this work, we have confirmed that, in a more general way,
there is no formation of gravastar even with the introduction of
a $L_e \ne \infty$.

On the other hand, for $m \ne 0$, if $L_i=L_e$ ($\alpha=1$) we have formation
of black hole or stable gravastar.  
These gravastars are only possible for $1 < \gamma \le 1.5$, 
satisfying all the dominant energy conditions. 
It is interesting to remark that this case can not be compared to 
other one already studied by us \cite{JCAP1}, except for $L_i=L_e=\infty$, which 
was shown in the figures 6 and 7, in that paper, and in the figure 6 of this work. 
While we have gravastars there for $\gamma<1$, here the gravastar formation 
is limited to $1 < \gamma \le 1.5$, showing that these intervals are 
complementary to each other, except for $\gamma=1$.

In the general case,
i.e., $m \ne 0$, $L_i \ne L_e$, it was seen that there is a range of 
$L_{e}$ in which "bounded excursion" stable 
gravastars are found, i.e.,  $L_{e}^{min} < L_e < L_{e}^{max}$. 
(Reminding that the curve for $L_e=\infty$ is very close to the curve 
for $L_e=L_e^{max}$.) Stable gravastars were found for $L_e \ge L_e^{max}$. 
Besides, this interval depends on the values of $L_i$ and $\gamma$. 
Let us now compare figures 12 and 13 , presented here, 
with figures 8 and 10, from \cite{JCAP1}, respectively. We can 
state that, from figures 8 and 10 \cite{JCAP1}, the bigger is 
$L_i$ (for $L_e=\infty$) the bigger is the 
tendency to the collapse of the shell, forming a "bounded excursion" gravastar or a black hole. 
Moreover, from figures 9, 12, 13 and 14 of this paper, for a given $L_i$, 
the formation of gravastars depends on the value of $L_e$ 
($L_e > {L_e}^{min}$, with ${L_e}^{min}\geq L_i$) in a such way that, 
instead of what occurs for $L_i$, the smaller is $L_e$ the bigger is 
the tendency to the collapse.
These conclusions are in agreement to the gravastar requirement 
proposed by Horvat \& Ilijic \cite{grava1}.  
The reason is that the dark energy density inside the gravastar have 
to be greater than the surround spacetime, i.e., $L_i < L_e$.
All these results can be summarized in Table II. 

\clearpage

\myfigure{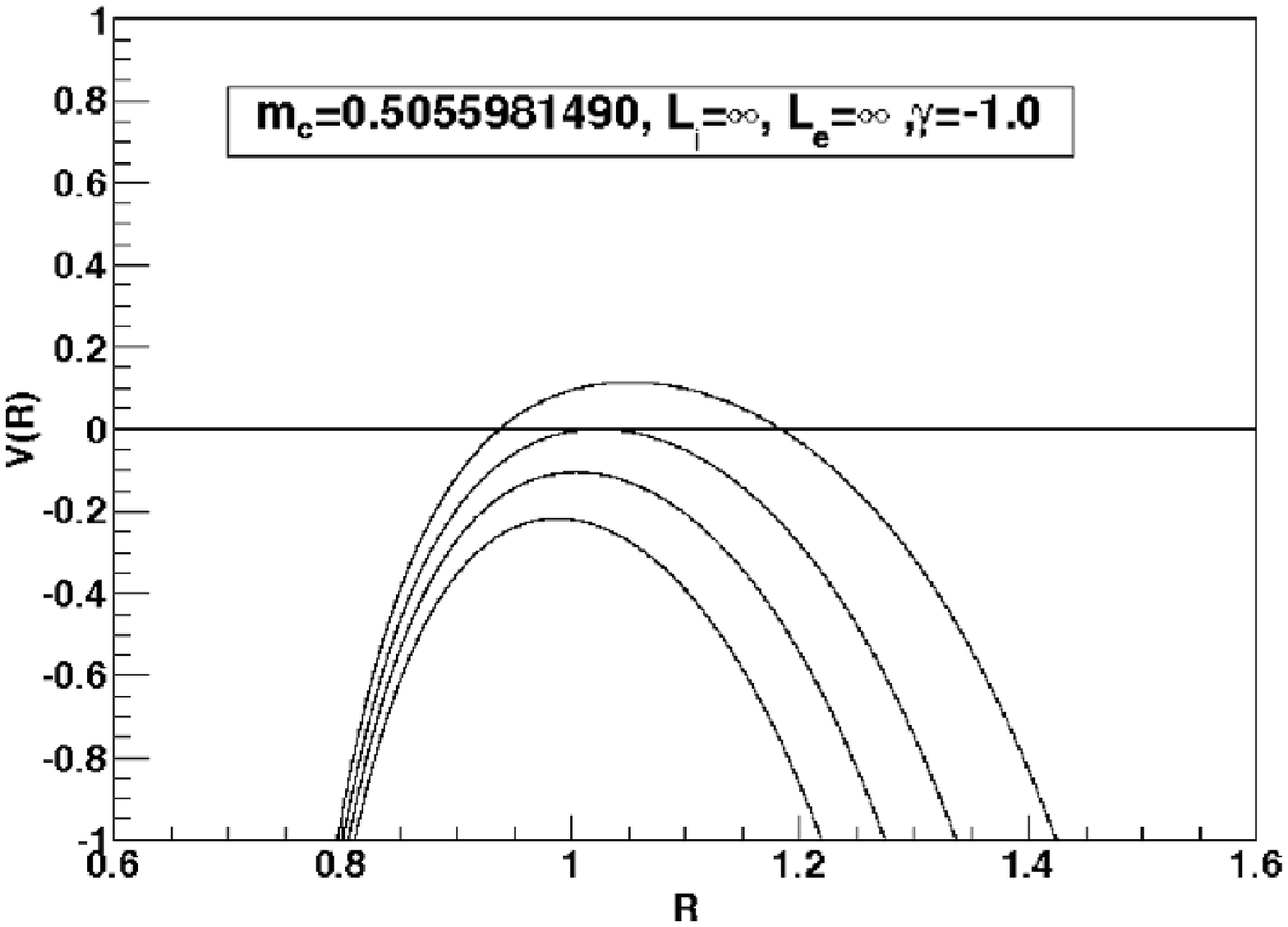}{The potential $V(R)$ for $\gamma=-1$,
$m_c=0.5055981490$, $L_e=\infty$, and  $L_i=\infty$ (the second curve
top-down). The others curves represent values for $m<m_c$ 
(first curve top-down) and $m>mc$ (the third and fourth curve top-down).
{\bf Case A}}
{figa}

\myfigure{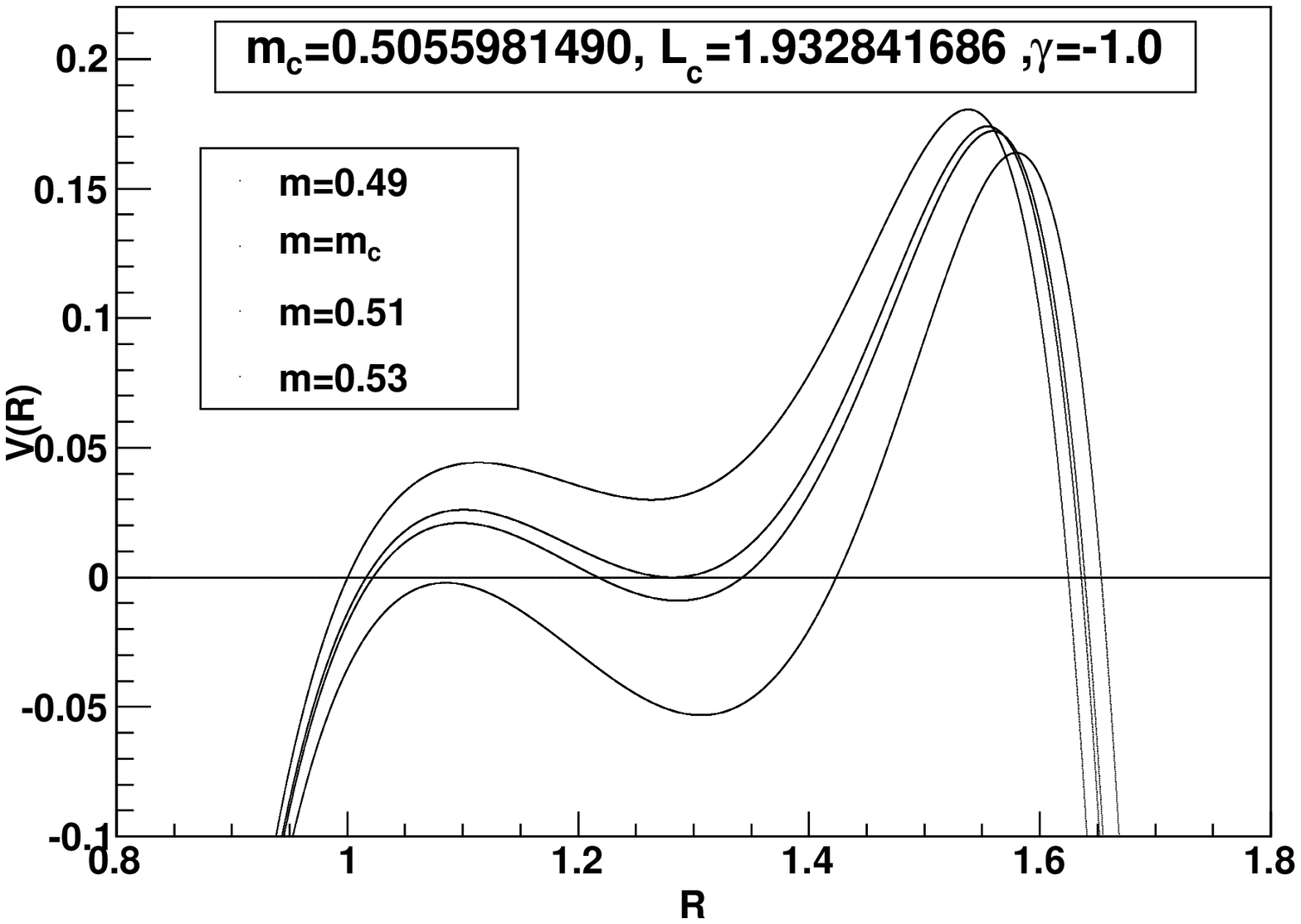}{The potential $V(R)$ for $\gamma=-1$,
$m_c=0.5055981490$, $L_i=L_c=1.932841686$ and $L_e=\infty$ (second curve
top-down). The first curve top-down assumes $m=0.49$. The third and
fourth curves top-down assume $m=0.51$ and $m=0.53$, respectively.
{\bf Case B}}
{figb}

\myfigure{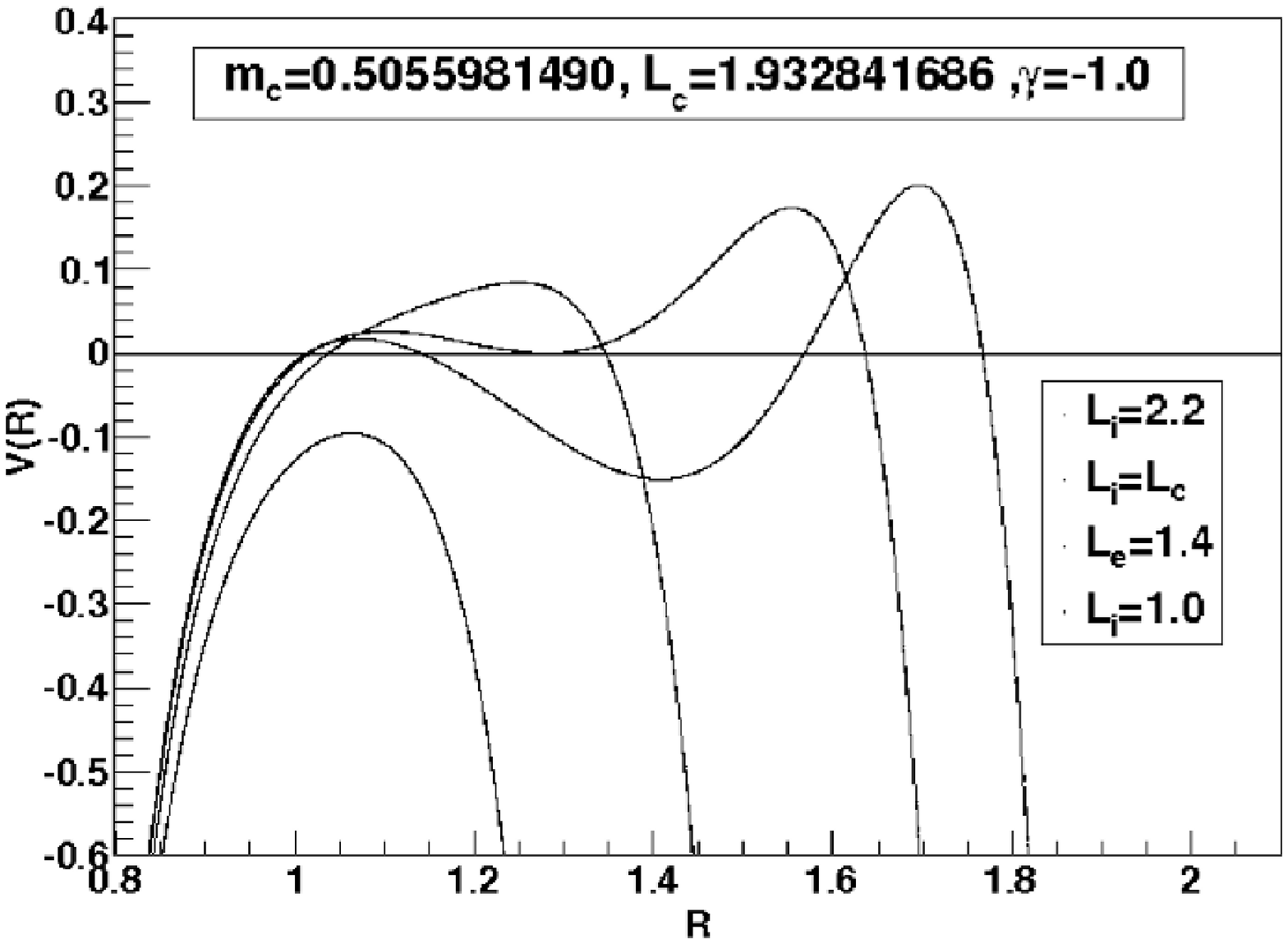}{The potential $V(R)$ for $\gamma=-1$,
$m_c=0.5055981490$, $L_i=1.932841686$ and $L_e=\infty$ (the second
curve top-down). The first curve assumes $L_i=1.4$.  The third and
fourth curves top-down assume $L_i=2.2$ and $L_i=1.0$, respectively.
{\bf Case B}}
{figc}

\myfigure{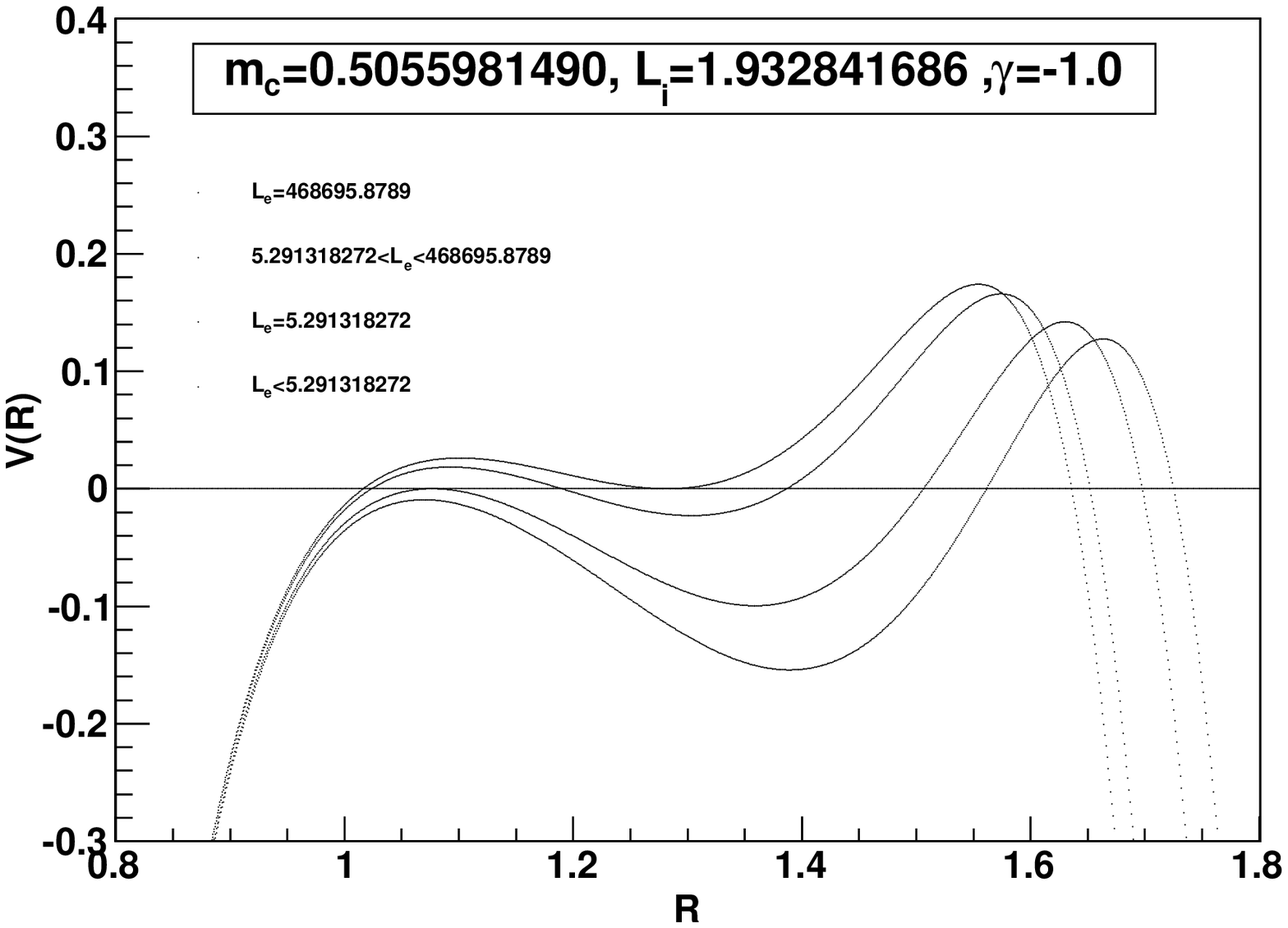}{The potential $V(R)$ for $\gamma=-1$,
$m_c=0.5055981490$, $L_i=1.932841686$ and $L_e=L_{e}^{max}=468695.8789$ 
(the first curve top-down). The second curve top-down is calculated
using $5.291318272<L_e<468695.8789$. The third curve
top-down is obtained assuming $L_e=5.291318272$.  The fourth curve
top-down assumes $L_e<5.291318272$. 
The curve for $L_e=\infty$ is very close
to the curve for $L_e=L_e^{max}$. 
{\bf Case B}}
{figd}

\myfigure{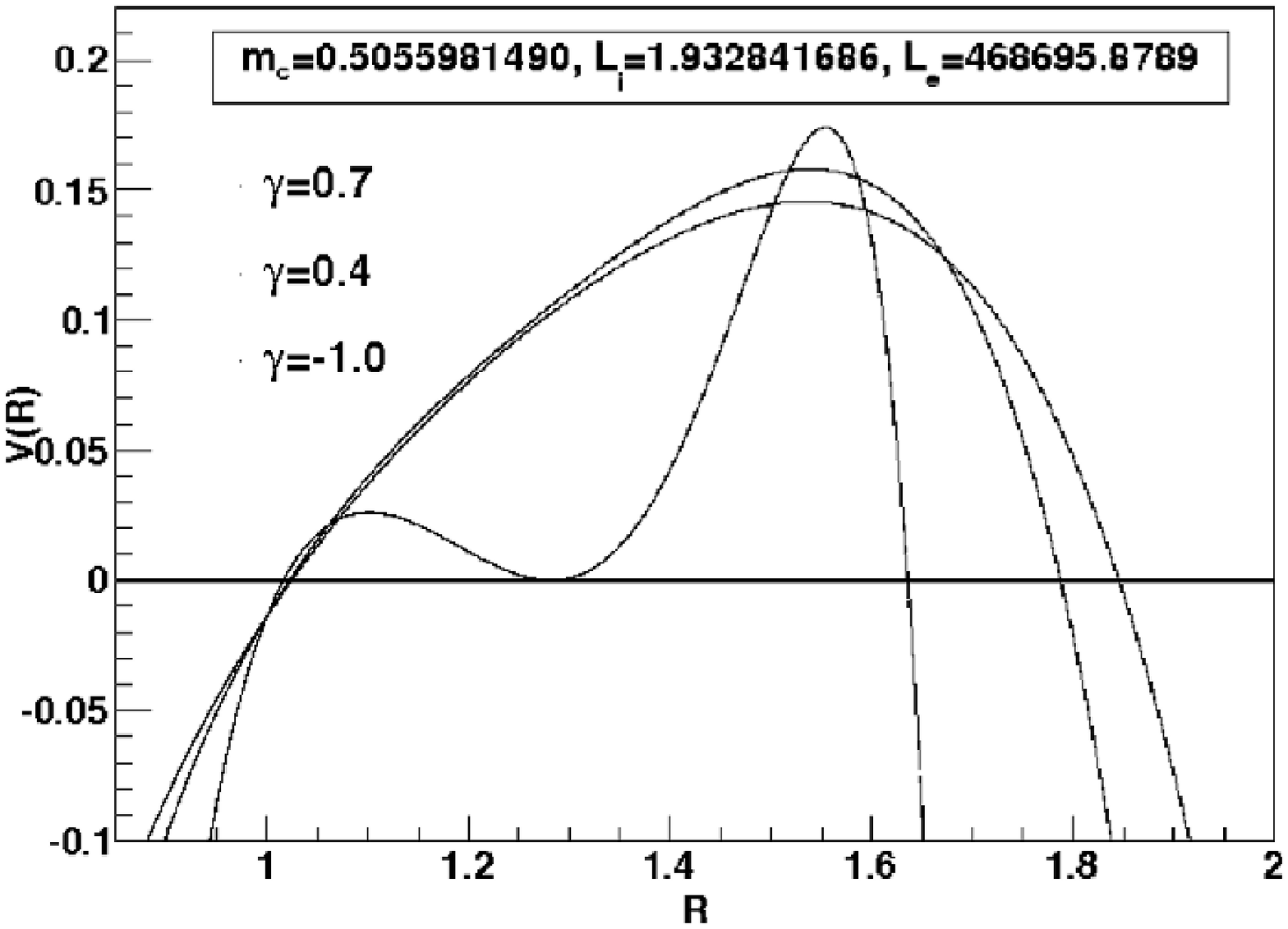}{The potential $V(R)$ for $\gamma=-1$,
$m_c=0.5055981490$, $L_i=1.932841686$ and $L_e=L_{e}^{max}=468695.8789$
(the curve that has a minimum). The others two curves top-down
use the values $\gamma=0.7$ and $\gamma=0.4$, respectively. 
{\bf Case B}}
{fige}

\myfigure{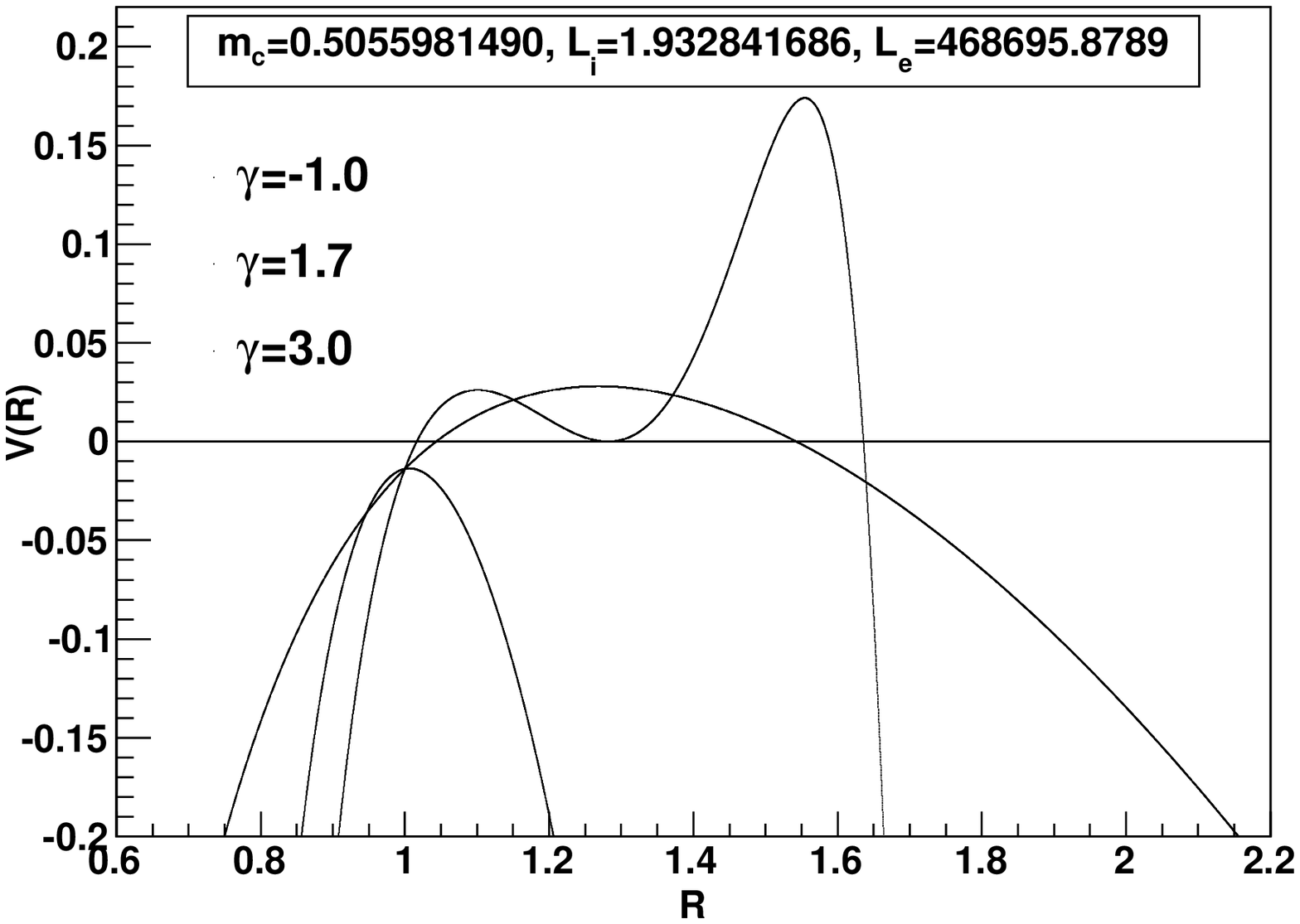}{The potential $V(R)$ for $\gamma=-1$,
$m_c=0.5055981490$, $L_i=1.932841686$ and $L_{e}^{max}=468695.8789$. 
(the curve that has a minimum). The others two curves top-down
use the values $\gamma=1.7$ and $\gamma=3$, respectively. 
{\bf Cases B and D}}
{figf}

\myfigure{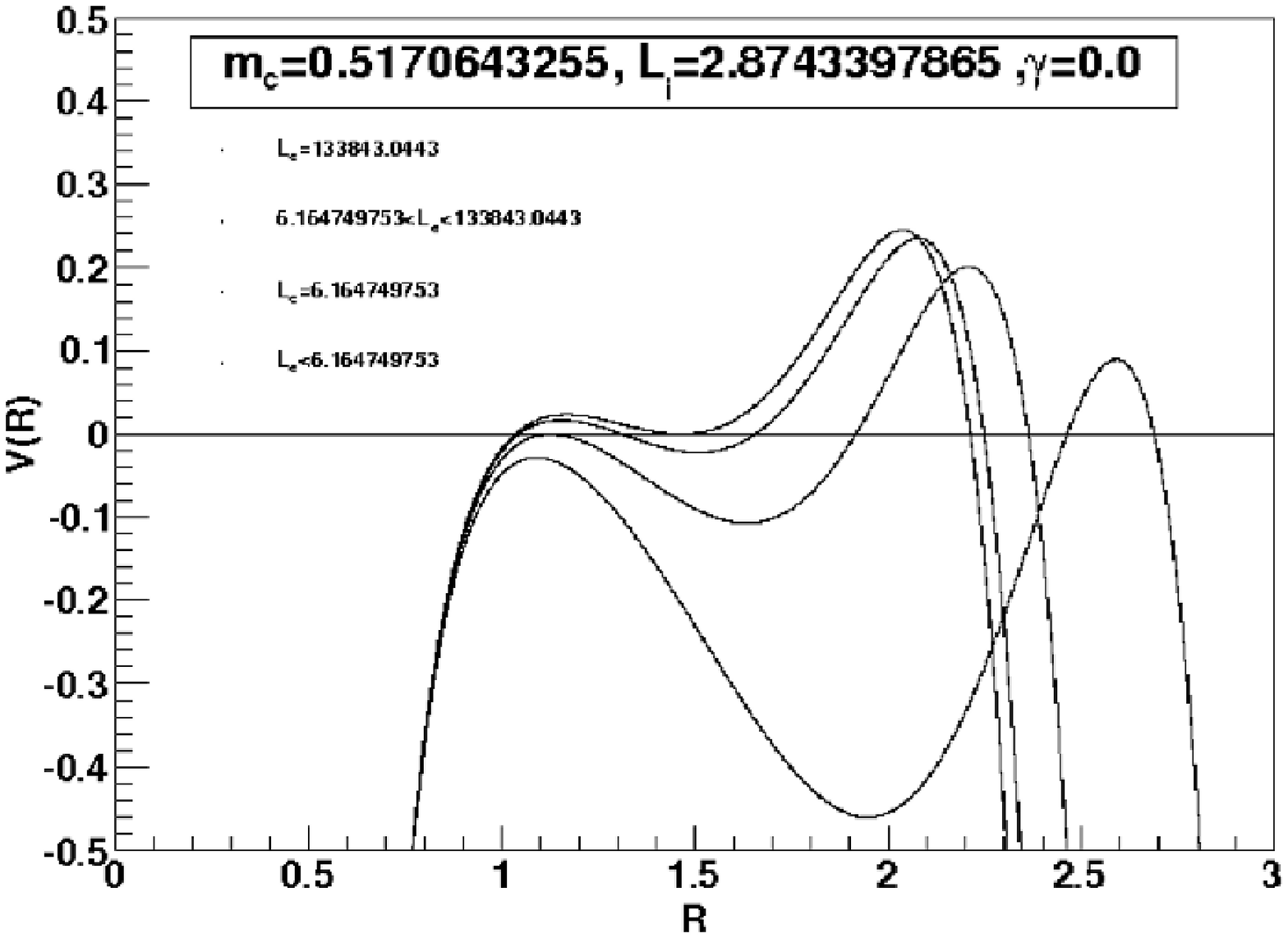}{The potential $V(R)$ for $\gamma=0$,
$L_e=133843.0443$, $L_i=2.8743397865$ and $m_c=0.5170643255$
(the first curve top-down). The second curve top-down is calculated
using $6.16479753<L_e<133843.0443$. The third curve
top-down is obtained assuming $L_e=6.16479753$.  The fourth curve
top-down assumes $L_e<6.164749753$. 
The curve for $L_e=\infty$ is very close
to the curve for $L_e=L_e^{max}$. 
These curves generalize the results presented in the 
figure 8 from \cite{JCAP1}. {\bf Case B}}
{fig1}

\myfigure{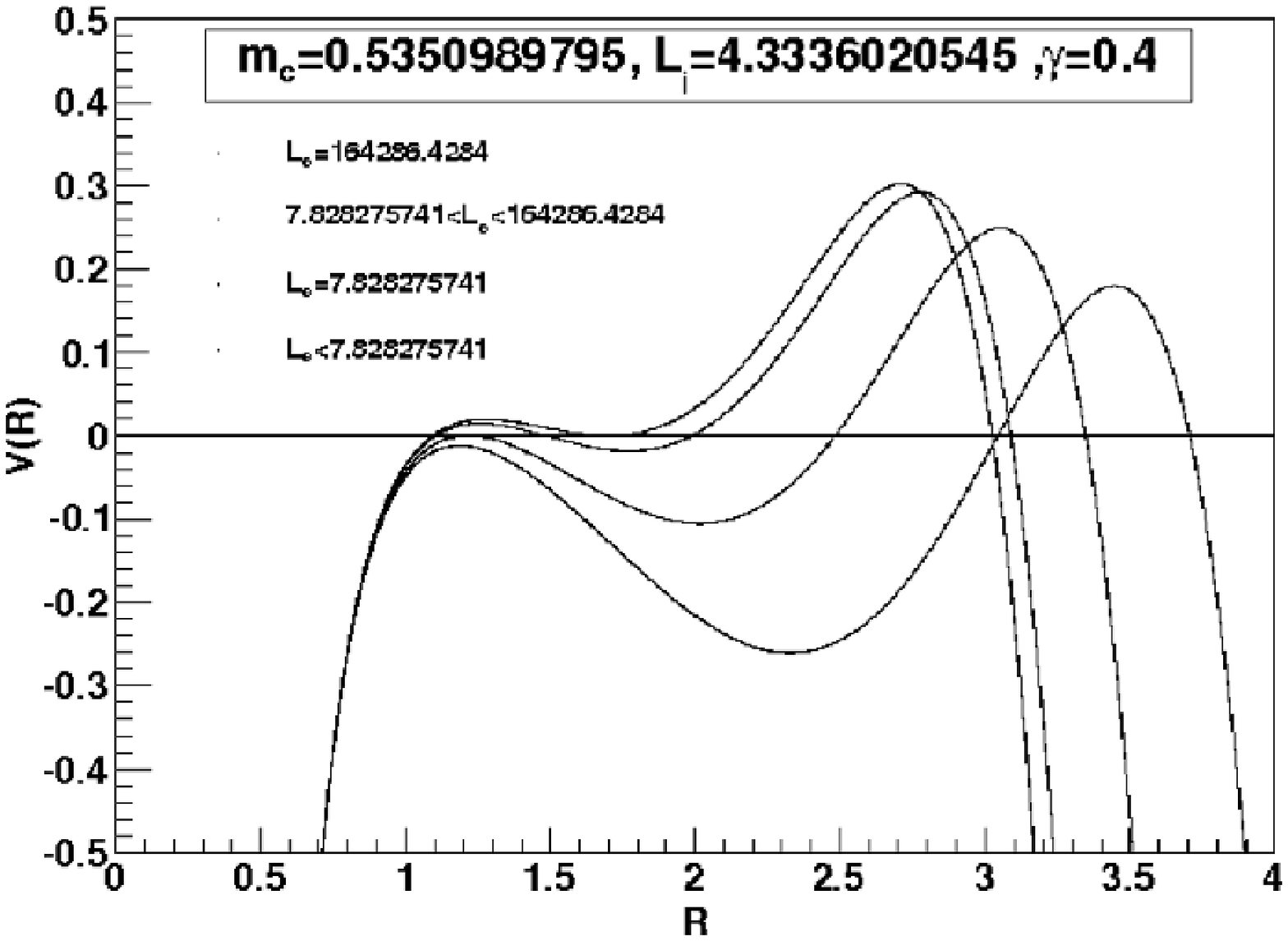}{The potential $V(R)$ for $\gamma=0.4$,
$L_e=164286.4284$, $L_i=4.3336020545$ and $m_c=0.5350989795$ 
(the first curve top-down). The second curve top-down is calculated 
using $7.828275741<L_e<164286.4284$. The third curve
top-down is obtained assuming $L_e=7.828275741$.  The fourth curve
top-down assumes $L_e<7.828275741$. 
The curve for $L_e=\infty$ is very close
to the curve for $L_e=L_e^{max}$. 
These curves generalize the results presented in the 
figure 10 from \cite{JCAP1}. {\bf Case B}}
{fig2}

\myfigure{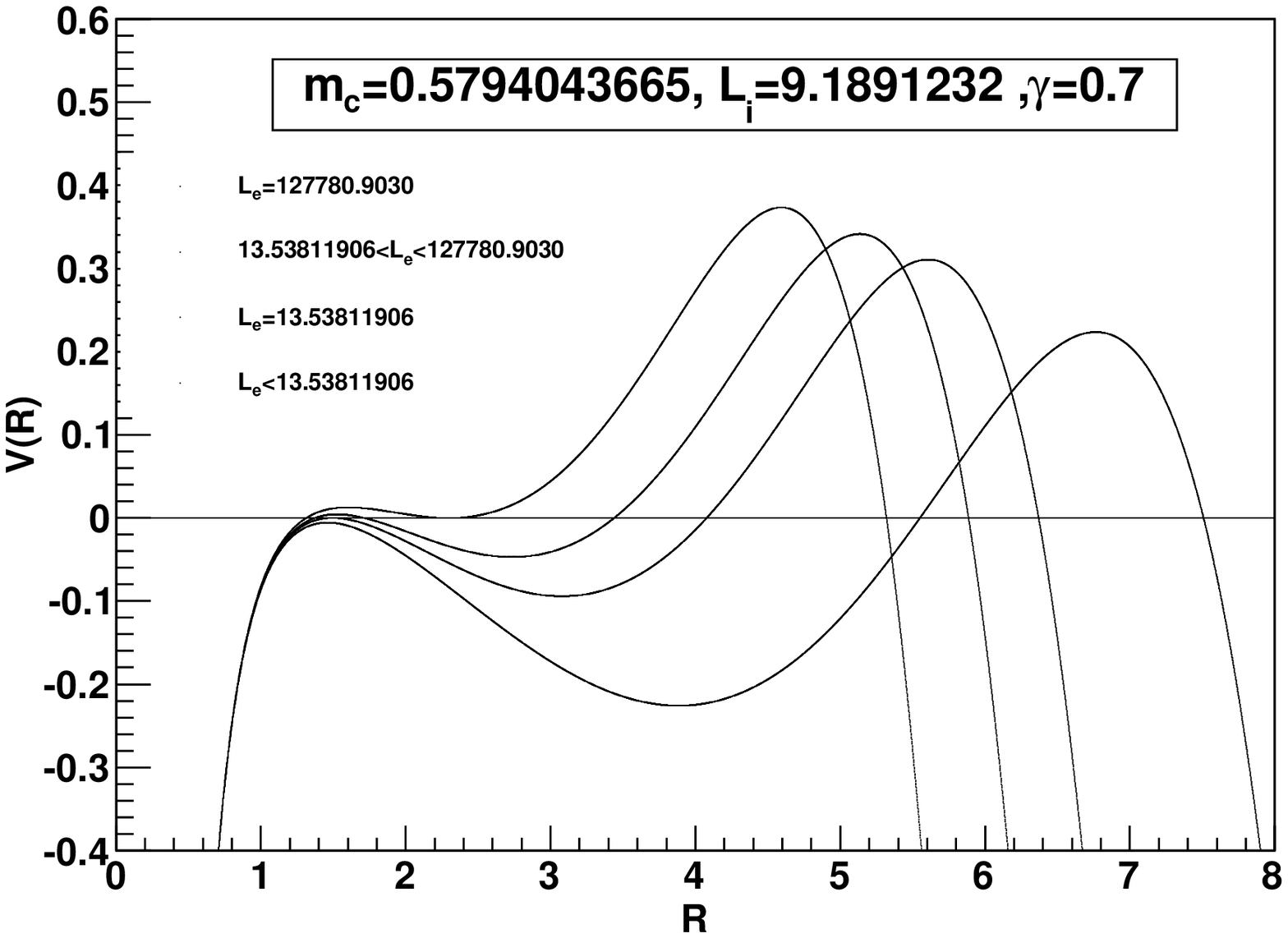}{The potential $V(R)$ for $\gamma=0.7$,
$L_e=127780.9030$, $L_i=9.1891232$ and $m_c=0.5794043665$ 
(the first curve top-down). The second curve top-down is calculated 
using $13.53811906<L_e<127780.9030$. The third curve
top-down is obtained assuming $L_e=13.53811906$.  The fourth curve
top-down assumes $L_e<13.53811906$. 
The curve for $L_e=\infty$ is very close
to the curve for $L_e=L_e^{max}$. 
These curves generalize the results presented in the 
figure 12 from \cite{JCAP1}. {\bf Case B}}
{fig3}

\myfigure{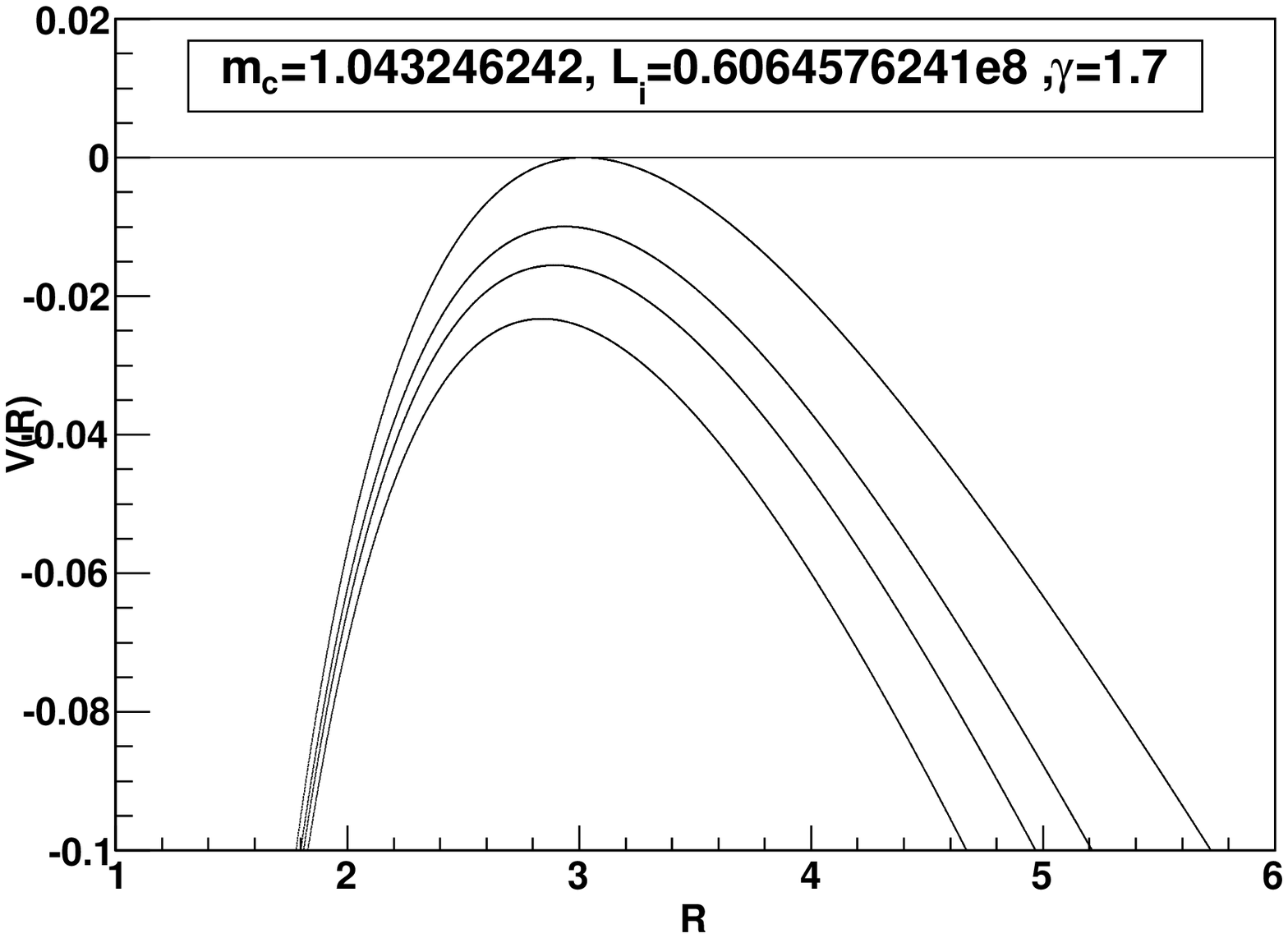}{The potential $V(R)$ for $\gamma=1.7$,
$L_i=0.6064576241\times 10^8$ and $m_c=1.043246242$ 
(the first curve top-down). The others curves represent $m>m_c$. 
The potential is insensible for variations of $L_e$. 
These curves generalize the results presented in the figure 20 
from \cite{JCAP1}.
{\bf Case C}}
{fig4}

\myfigure{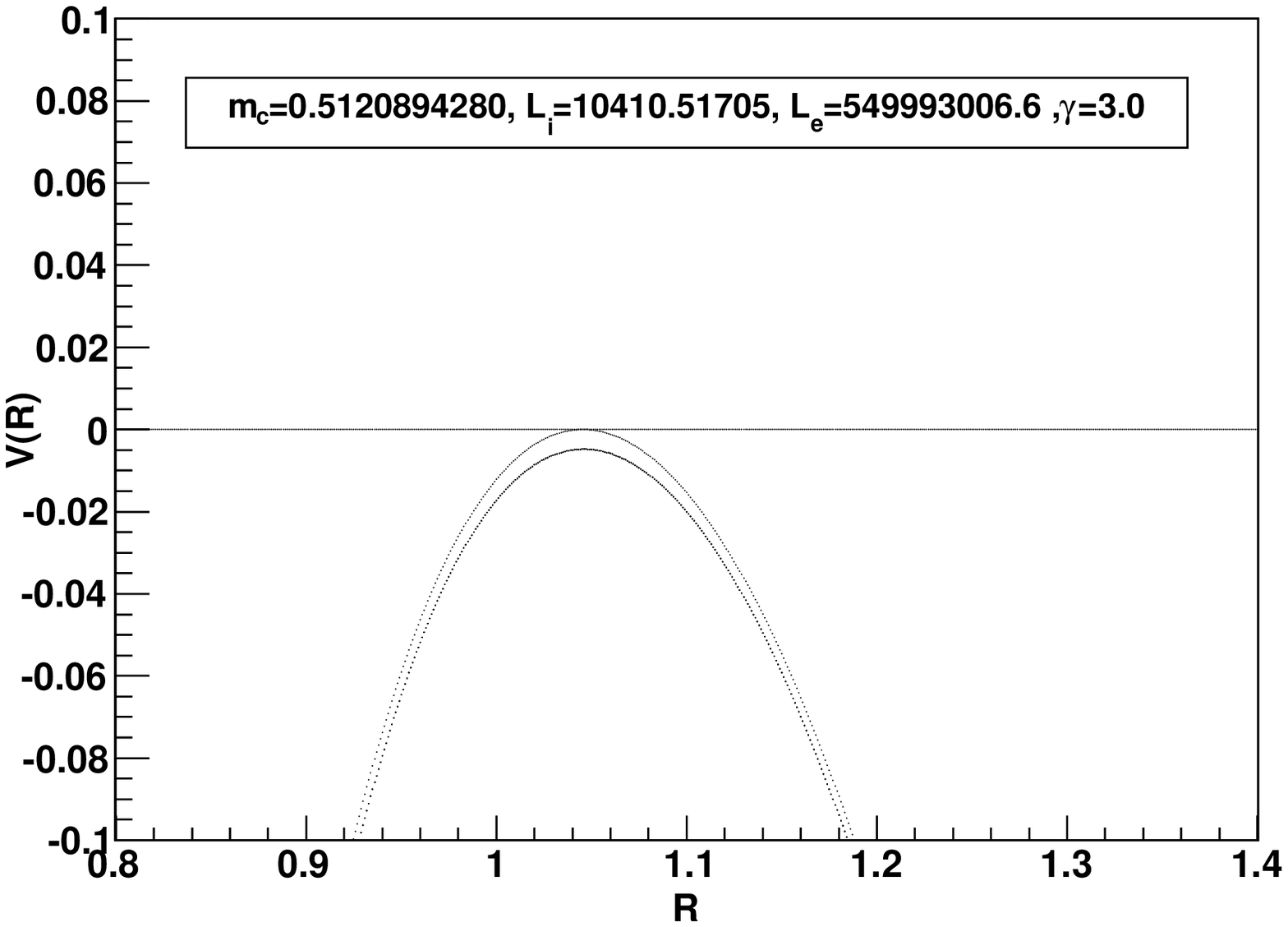}{The potential $V(R)$ for $\gamma=3$,
$L_e=549993006.6$, $L_i=10410.51705$ and $m_c=0.5120894280$
(the first curve top-down). The second curve represent $m>m_c$. 
{\bf Case D}}
{fig5}

\myfigure{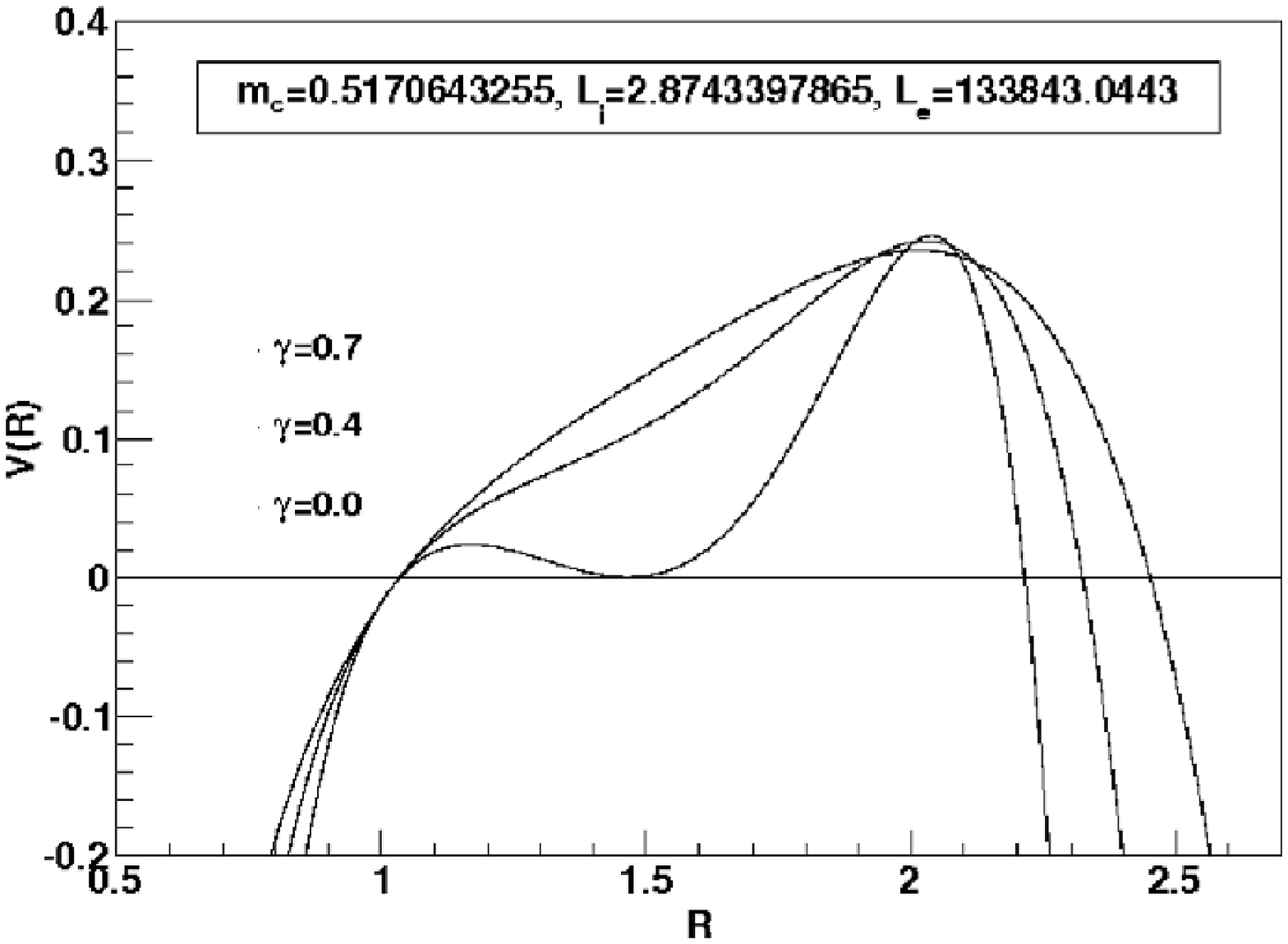}{The potential $V(R)$ for $\gamma=0.7$,
$L_e=133843.0443$, $L_i=2.8743397865$ and $m_c=0.5170643255$ 
(the first curve top-down). The second curve top-down is calculated 
using $\gamma=0.4$. The third curve
top-down is obtained assuming $\gamma=0$.
{\bf Case B}}
{fig6}

\myfigure{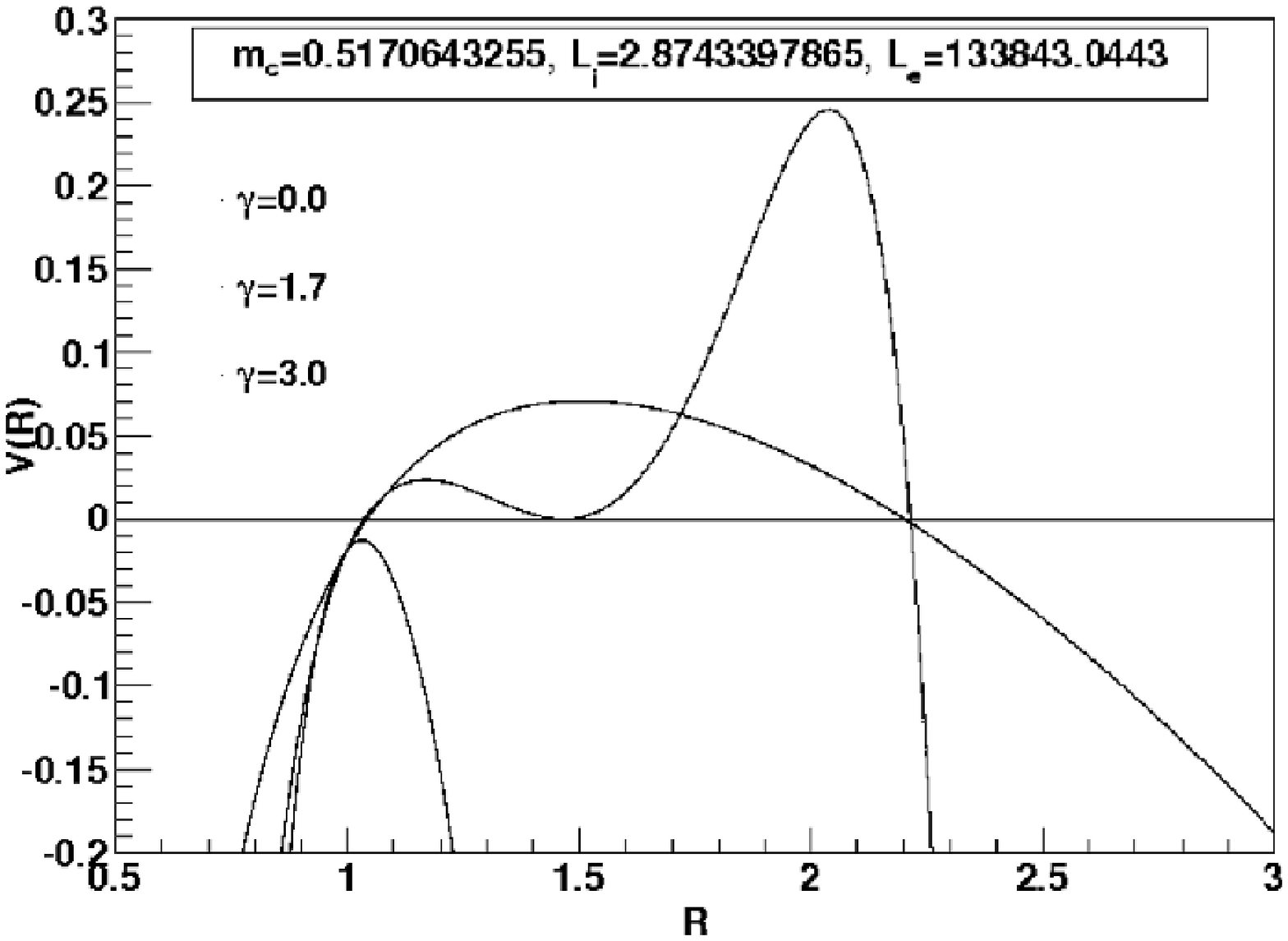}{The potential $V(R)$ for $\gamma=0$,
$L_e=133843.0443$, $L_i=2.8743397865$ and $m_c=0.5170643255$ 
(the first curve top-down). The second curve top-down is calculated 
using $\gamma=1.7$. The third curve
top-down is obtained assuming $\gamma=3$.
{\bf Case D}}
{fig7}

\myfigure{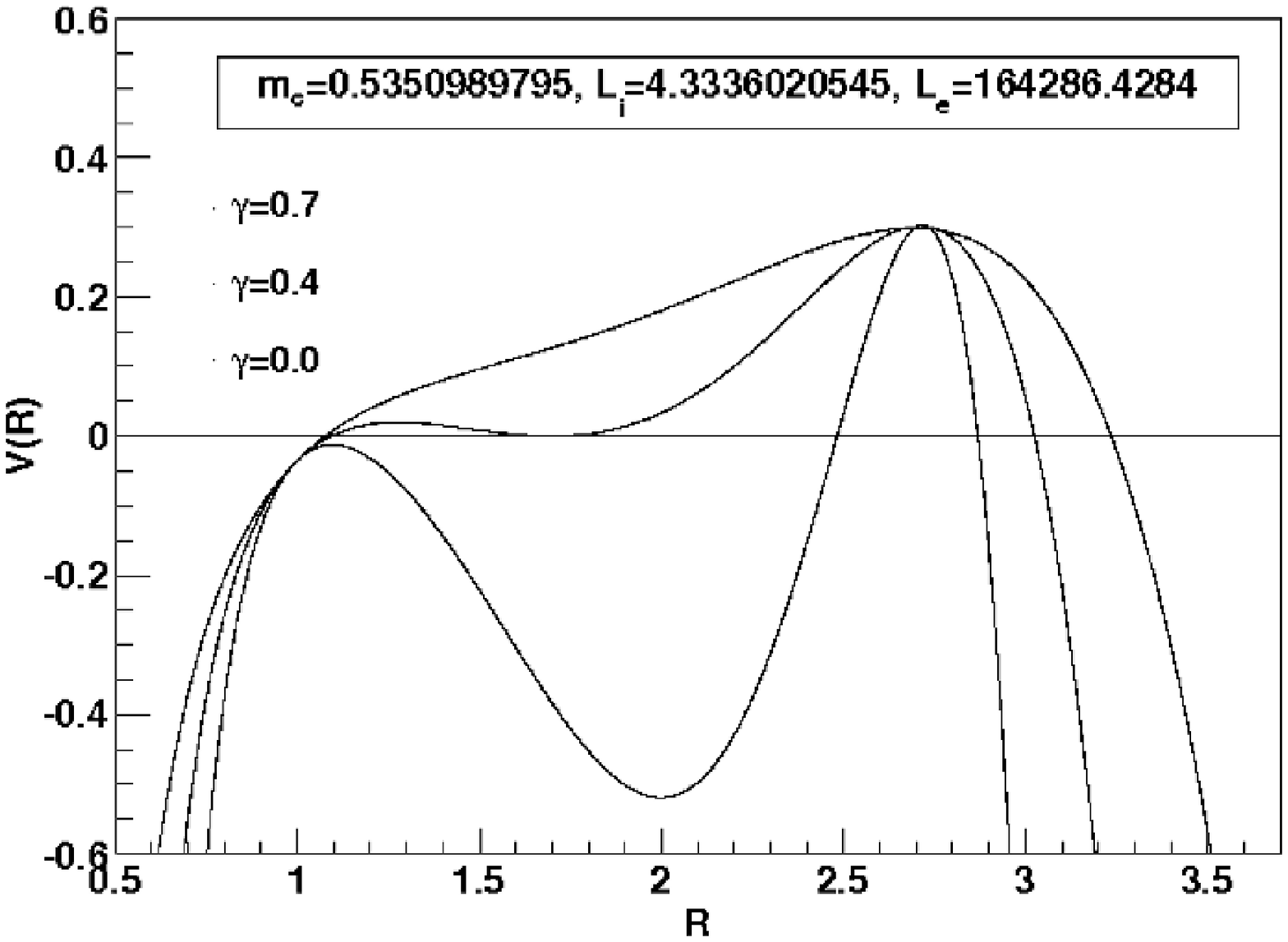}{The potential $V(R)$ for $\gamma=0.7$,
$L_e=164286.4284$, $L_i=4.3336020545$ and $m_c=0.5350989795$ 
(the first curve top-down). The second curve top-down is calculated 
using $\gamma=0.4$. The third curve
top-down is obtained assuming $\gamma=0$.
{\bf Case B}}
{fig8}

\myfigure{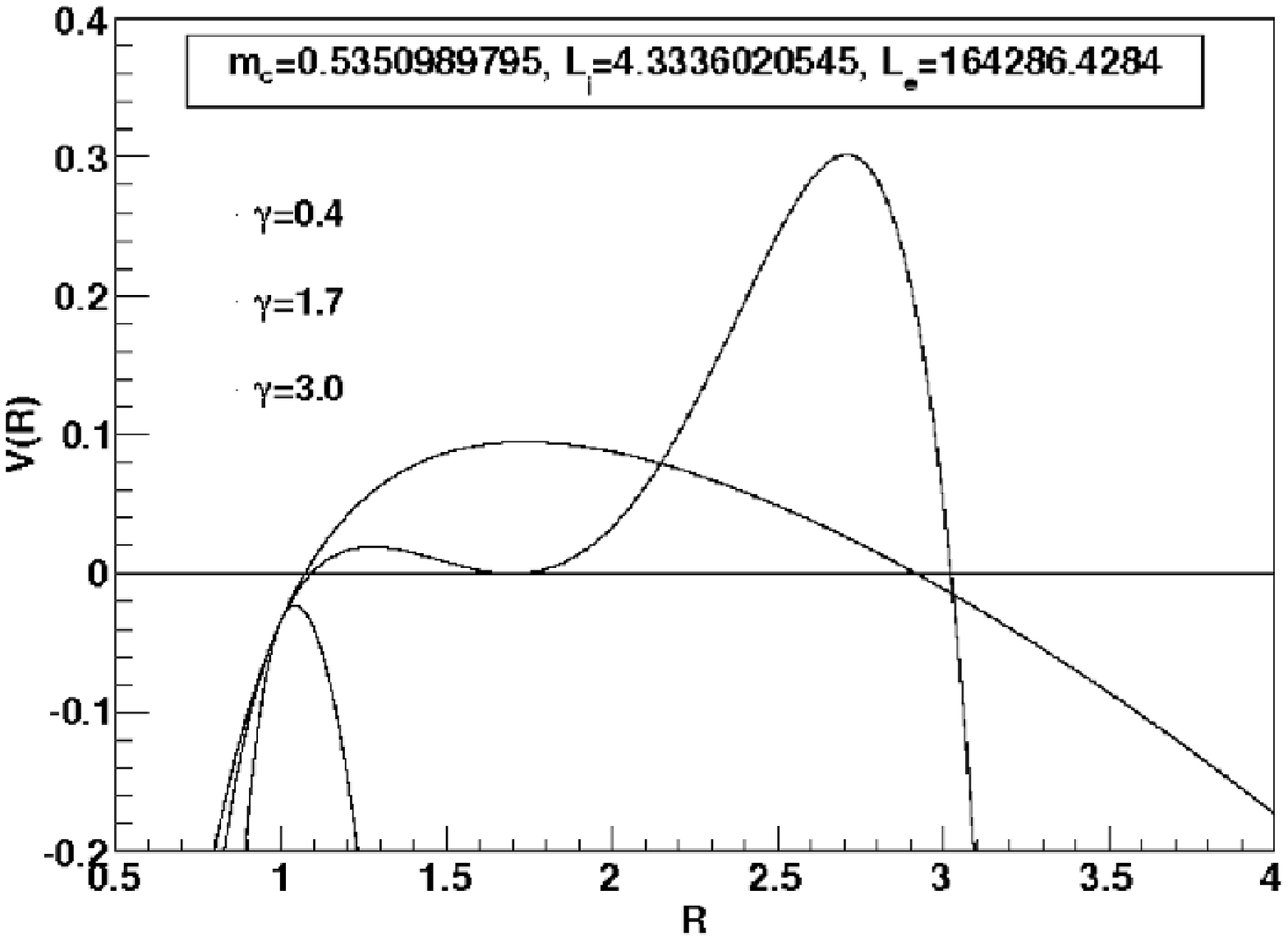}{The potential $V(R)$ for $\gamma=0.4$,
$L_e=164286.4284$, $L_i=4.3336020545$ and $m_c=0.5350989795$ 
(the first curve top-down). The second curve top-down is calculated 
using $\gamma=1.7$. The third curve
top-down is obtained assuming $\gamma=3$.
{\bf Case D}}
{fig9}

\myfigure{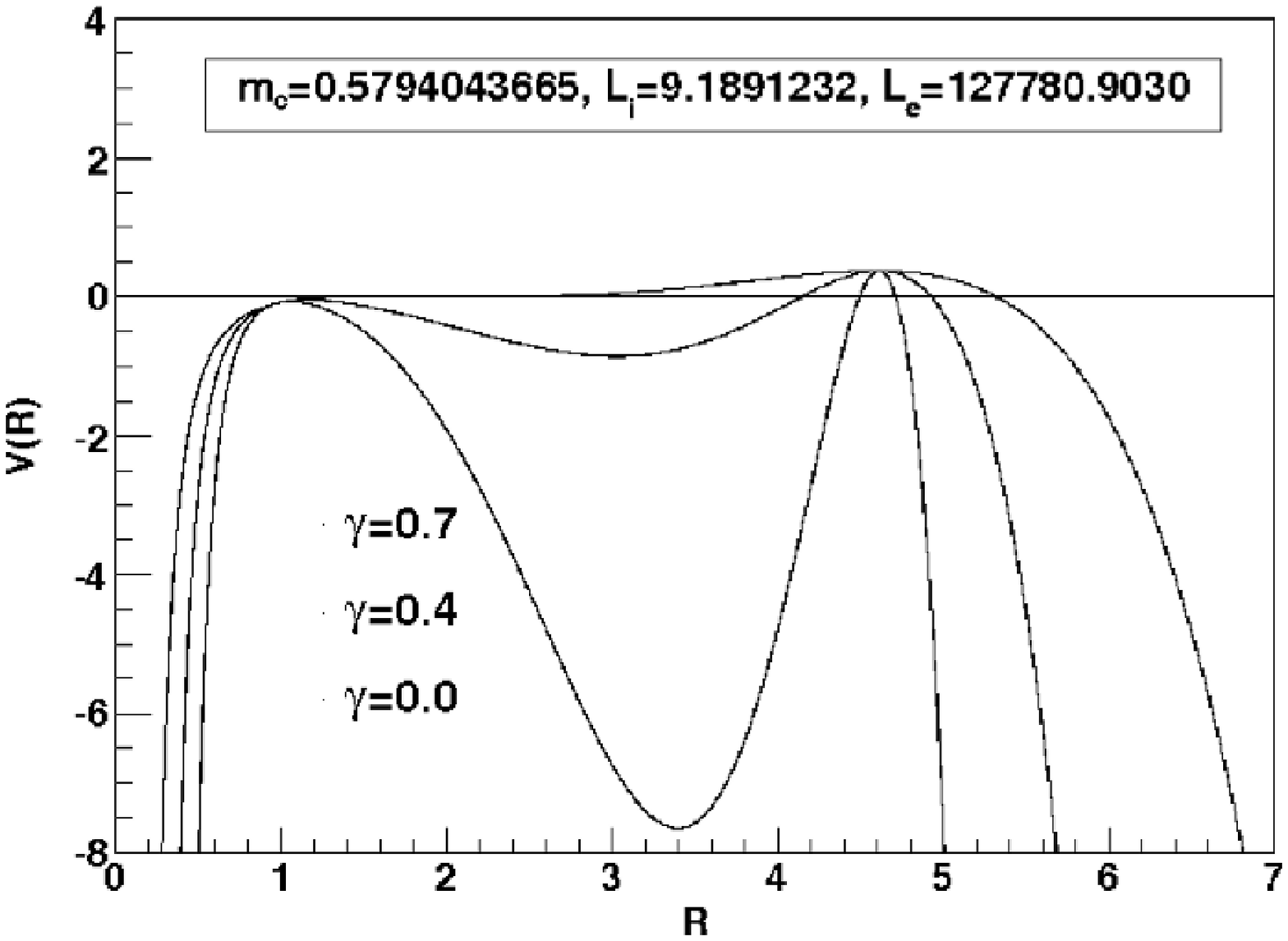}{The potential $V(R)$ for $\gamma=0.7$,
$L_e=127780.9030$, $L_i=9.1891232$ and $m_c=0.5794043665$ 
(the first curve top-down). The second curve top-down is calculated 
using $\gamma=0.4$. The third curve
top-down is obtained assuming $\gamma=0$.
{\bf Case B}}
{fig10}

\myfigure{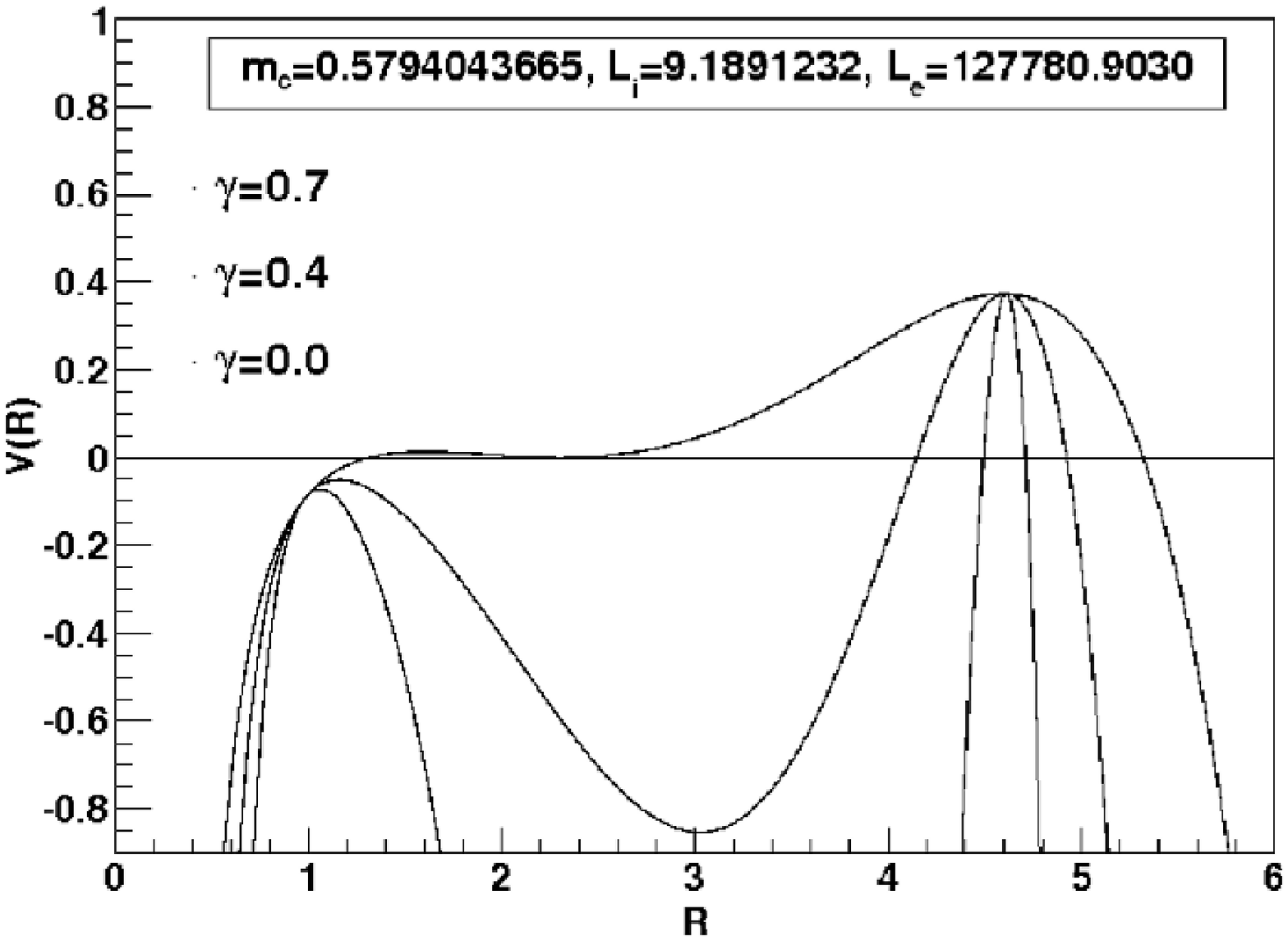}{The potential $V(R)$ for $\gamma=0.7$,
$L_e=127780.9030$, $L_i=9.1891232$ and $m_c=0.5794043665$ 
(the first curve top-down). The second curve top-down is calculated 
using $\gamma=0.4$. The third curve
top-down is obtained assuming $\gamma=0$.
{\bf Case B}}
{fig11}

\myfigure{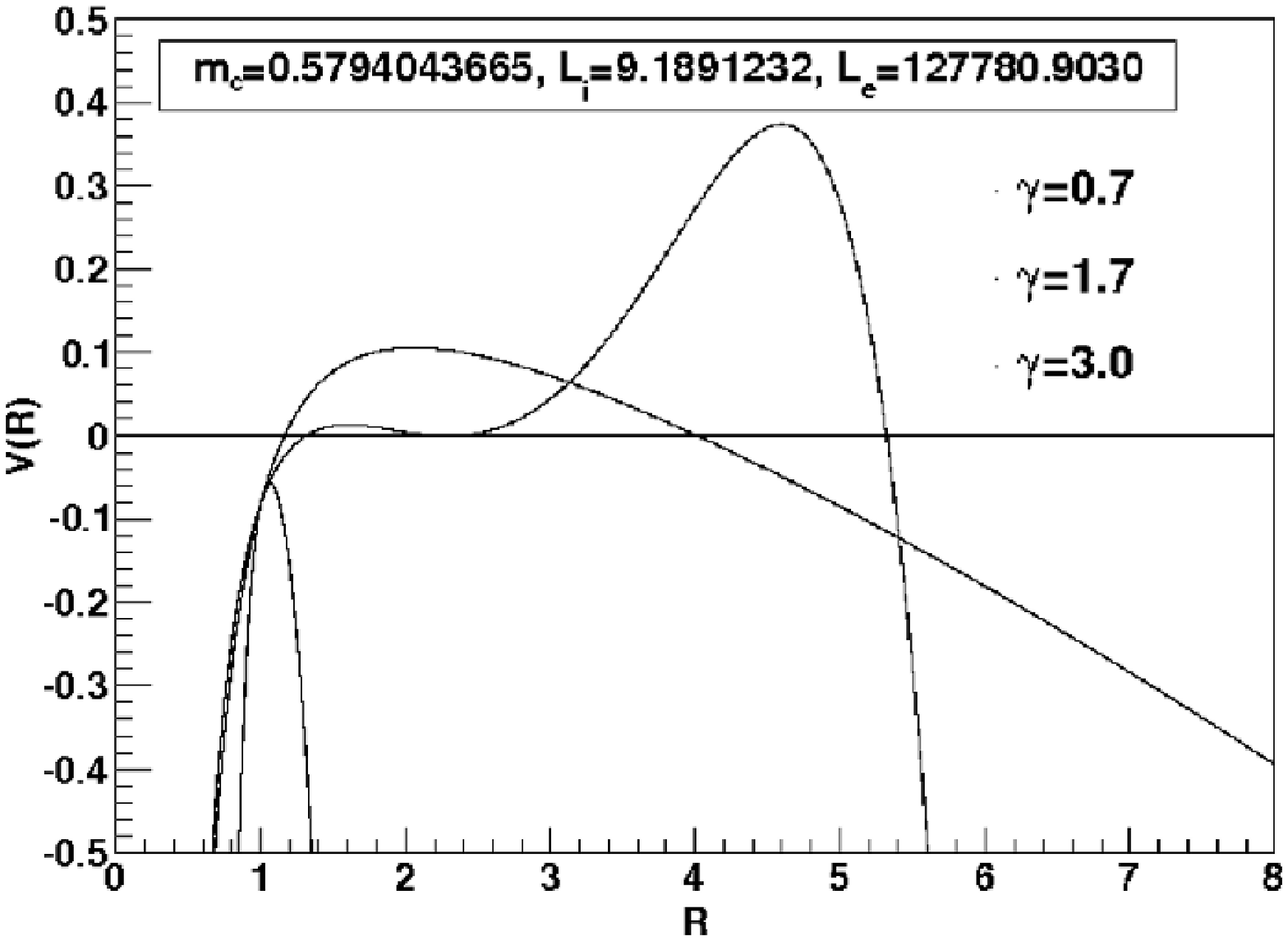}{The potential $V(R)$ for $\gamma=0.7$,
$L_e=127780.9030$, $L_i=9.1891232$ and $m_c=0.5794043665$ 
(the first curve top-down). The second curve top-down is calculated 
using $\gamma=1.7$. The third curve
top-down is obtained assuming $\gamma=3$.
{\bf Case D}}
{fig12}

\begin{acknowledgments}
We thank Dr. Anzhong Wang for helpful discussions that improved this work.
The financial assistance from 
FAPERJ/UERJ (MFAdaS) is gratefully acknowledged. The
author (RC) acknowledges the financial support from FAPERJ (no.
E-26/171.754/2000, E-26/171.533/2002 and E-26/170.951/2006). 
The authors (RC and MFAdaS) also acknowledge the financial support from 
Conselho Nacional de Desenvolvimento Cient\'{\i}fico e Tecnol\'ogico - 
CNPq - Brazil.  The author (MFAdaS) also acknowledges the financial support
from Financiadora de Estudos e Projetos - FINEP - Brazil (Ref. 2399/03).
\end{acknowledgments}

\end{document}